\documentclass[fleqn,usenatbib]{mnras}

\usepackage{newtxtext,newtxmath}

\usepackage[T1]{fontenc}

\DeclareRobustCommand{\VAN}[3]{#2}
\let\VANthebibliography\thebibliography
\def\thebibliography{\DeclareRobustCommand{\VAN}[3]{##3}\VANthebibliography}

\usepackage{graphicx}	
\usepackage{amsmath}	
\usepackage{comment}
\usepackage{siunitx}
\usepackage{caption}
\usepackage{subfig}
\usepackage{multirow}
\usepackage{rotating}
\usepackage{booktabs}
\usepackage{enumitem}
\usepackage{arydshln}

\newcommand{\pcmmm}{\,cm$^{-3}$}	
\newcommand{\kms}{km\,s$^{-1}$} 

\newcommand{\WHz}{W\,Hz$^{-1}$}
\newcommand{\pc}{\,pc$^{-2}$}

\newcommand{\fw}{$f_\mathrm{W}$}
\newcommand{\lclump}{$L_\mathrm{clump}$}
\newcommand{\lradio}{$L_\mathrm{radio}$}
\newcommand{\mclump}{$M_\mathrm{clump}$}
\newcommand{\mcomplex}{$M_\mathrm{complex}$}
\newcommand{\mgmc}{$M_\mathrm{GMC}$}
\newcommand{\lm}{\lradio/\mcomplex}

\title[\ion{H}{ii} regions and supernova remnants associated with molecular clouds]{\ion{H}{ii} regions and supernova remnants associated with molecular clouds: A pilot study with the SARAO MeerKAT Galactic Plane Survey}

\author[M. O. Langa et al.]{
Moses O. Langa,$^{1,2}$\thanks{E-mail: langamoses1@gmail.com}
Mark A. Thompson,$^{2}$
Andrew J. Rigby,$^{2}$
Gwenllian M. Williams,$^{3}$
Mubela Mutale,$^{2}$
\newauthor
Paul O. Baki,$^{1}$
James O. Chibueze,$^{4,5}$
and Willice O. Obonyo$^{1,4}$
\\
$^{1}$Department of Astronomy and Space Science, Technical University of Kenya, Nairobi, P.O. Box 52428-00200, Kenya\\
$^{2}$School of Physics and Astronomy, University of Leeds, Leeds LS2 9JT, UK\\
$^{3}$Department of Physics, Aberystwyth University, Ceredigion, Cymru, SY23 3BZ, UK\\
$^{4}$Department of Mathematical Sciences, University of South Africa, Cnr
Christian de Wet Rd and Pioneer Avenue, Florida Park, 1709, Roodepoort,
South Africa\\
$^{5}$Department of Physics and Astronomy, Faculty of Physical Sciences, University of Nigeria, Carver Building, 1 University Road, Nsukka 410001, Nigeria\\
}

\date{Accepted 2025 November 13. Received 2025 November 06; in original form 2025 January 23}

\pubyear{\the\year{}}

\begin{document}
\label{firstpage}
\pagerange{\pageref{firstpage}--\pageref{lastpage}}
\maketitle

\begin{abstract}
Massive stars (mass$>8\mathrm{M_\odot}$) release vast amounts of energy into the interstellar medium through their stellar winds, photoionising radiation and supernova explosions. These processes may compress nearby regions, triggering further star formation, but the significance of triggered star formation across the Galactic disc is not well understood. This pilot study combines 1.3\,GHz continuum data from the South African Radio Astronomy Observatory (SARAO) MeerKAT Galactic Plane Survey (SMGPS) with $^{13}$CO (2--1) data from the Structure, Excitation, and Dynamics of the Inner Galactic Interstellar Medium (SEDIGISM) survey to identify and examine molecular clouds associated with \ion{H}{ii} regions and supernovae remnants (SNRs). We focus on their physical properties and massive star formation potential. We identify 268 molecular clouds from the SEDIGISM tile covering the Galactic plane region $341^{\circ} \leq \ell \leq 343^{\circ}$ and $|b|\leq 0.5^{\circ}$, of which 90 clouds (34 per cent) are associated with SMGPS extended sources. Compared to unassociated clouds, we find that associated clouds exhibit significantly higher mean mass ($\sim$ 9600 M$_\odot$ vs. $\sim$ 2500 M$_\odot$) and average gas surface density ($\sim$ 104 M$_\odot$\pc vs. $\sim$ 67 M$_\odot$\pc), and slightly elevated but comparable virial parameters. We also find that the size-linewidth scaling relation is steeper for associated clouds compared to unassociated clouds. In addition, radio luminosity shows a positive correlation with total complex mass, and the ratio \lm\ increases with source size, consistent with an evolutionary sequence where expanding \ion{H}{ii} regions progressively disrupt their natal molecular environment. These findings suggest an enhanced dynamical activity for the associated clouds and support the hypothesis that feedback from massive stars influences molecular cloud properties and may trigger star formation.
\end{abstract}




\section{Introduction}
Star formation occurs within the densest ($10^{3}$--$10^{6}$\,\pcmmm) and coldest (10--20\,K) regions of molecular clouds \citep[][]{2012ARA&A..Kennicutt,2017A&A...Volschow}. This happens when large clouds of molecular hydrogen become unstable against their own gravity and collapse, hence new stars are born. It has long been suspected that, in some instances, this collapse may be triggered by external environmental factors, such as the intense ionizing radiation produced by massive stars or enormous shock waves caused when massive stars explode at the end of their lives. The energy produced by these stars compresses the nearby interstellar medium, driving shocks that may induce gravitational collapse and subsequently triggering the formation of a new generation of stars through a process known as radiation-driven implosion (RDI) \citep{1977ApJ...Elmegreen, 1989ApJ....Bertoldi, 1994A&A....Lefloch}, and possibly setting off a cycle of propagating massive star formation throughout giant molecular clouds \citep[e.g.][]{2005A&A...433..565D}. Massive star formation is also linked to other physical processes, such as the `collect and collapse' process \citep[][]{1998ASPC..148..150E, 2005A&A...433..565D, 2006A&A...446..171Z, 2008ASPC..Deharveng} which is characterised by the gradual expansion of \ion{H}{ii} regions into the ISM, which accumulate gas and dust within the boundaries of ionization and shock fronts. Eventually, the surrounding layer becomes gravitationally unstable and fragments into new cores, some of which might be massive enough to produce another generation of massive stars \citep[][]{1994A&A...290..421W, 2006A&A...446..171Z, 2009A&A...496..177D, 2010A&A...523A...6D}. The missing piece of the puzzle is a well selected and statistically significant sample of molecular clouds that have undergone shocks or interaction with nearby \ion{H}{ii} regions or supernovae.

\ion{H}{ii} regions and supernova remnants (SNRs) are critical for studying massive star formation, as they are exclusively associated with massive stars in regions of ongoing stellar activity \citep{2007ARA&A..Zinnecker}. At radio wavelengths, \ion{H}{ii} regions are characterised by thermal free–free emission from ionized gas \citep{1967ApJ...147..Mezger, 2009tra..book.....Wilson}, which is frequently coincident with mid-infrared dust and PAH (polycyclic aromatic hydrocarbon) emission tracing their ionization fronts and bubble rims \citep{2005pcim.book...Tielens, 2008ApJ...Watson}. By contrast, SNRs typically exhibit non-thermal synchrotron emission with steep spectral indices, often appearing morphologically distinct from thermal \ion{H}{ii} regions \citep{2008ARA&A..Reynolds, 2019JApA...Green}. In some cases, SNR emission overlaps with shocked molecular gas, revealed through broadened CO line profiles \citep{1998ApJ...Frail, 2010ApJ...Jiang} or maser activity \citep{1994ApJ...Frail, 2009ApJ...Hewitt}, providing direct evidence of energetic feedback. Therefore, these complementary diagnostics across the radio and infrared regimes are essential for distinguishing between \ion{H}{ii} regions and SNRs and for interpreting their spatial and physical relationship with nearby molecular clouds \citep{2011piim.book...Draine, 2014ApJS..Anderson(2)}.

A number of studies have been carried out, which examine \ion{H}{ii} regions across the radio-infrared (IR) regime and examine the interaction between \ion{H}{ii} and molecular cloud, especially those associated with bright-rimmed clouds \citep[e.g.,][]{1989ApJ...Sugitani, 1991ApJS...Sugitani, 1994ApJS...Sugitani, 2004A&A...Thompson(a), 2004A&A...Thompson(b), 2004A&A...Thompson(c), 2004A&A...Urquhart, 2005ApJ...Lee, 2006A&A...Urquhart, 2007A&A...Urquhart, 2008A&A...Morgan, 2008ASPC..Deharveng,2009MNRAS.Morgan, 2013ApJ...762...17C, 2022ApJ...Sharma, 2022PASA...Azatyan,2022AAS...Finley}, and/or supernova remnants \citep{2008MNRAS...Stupar, 2011MNRAS.414.Stupar, 2023ApJS...Zhou, 2024arXiv240916607A...Anderson} but the only large scale studies so far have been of infrared bubbles \citep[e.g.,][]{2012MNRAS.421..408T,2012ApJ...755...71K}. These infrared studies suggest that roughly 15 per cent of massive stars in the Milky Way may have been formed as a result of triggering processes from the expansion of \ion{H}{ii} regions, but are limited in the conclusions that can be drawn due to the lack of distance information and also in their exclusion of SNRs. There have been studies of clouds associated with \ion{H}{ii} regions and SNRs. For example, from their two surveys of molecular clouds around galactic SNRs, \cite{2002aprm.conf..Yamamoto} presented results that suggested that the SNRs are closely interacting with the molecular clouds and that supernova shocks can indeed compress nearby gas. \cite{2012A&A...Anderson} used radio and infrared observations to examine star formation around \ion{H}{ii} regions, with young stellar objects (YSOs) found near expanding ionized regions. \cite{2016ApJ...Kendrew} examined the properties of ATLASGAL molecular clumps near infrared bubbles using a two-point correlation analysis, and found a clear overdensity of massive cold clumps along bubble rims, with the effect becoming stronger for larger bubbles. Interestingly, the clumps with the highest column densities tend to remain toward bubble interiors, resisting displacement by the expanding edges. Spectroscopic ammonia observations further revealed that clumps near bubbles are denser, hotter, and more turbulent than those in the field, providing circumstantial evidence that these clumps are more likely to be massive star-forming sites. Recently, \cite{2023ApJS...Zhou} conducted a search for associations between molecular clouds and supernova remnants for nearly all SNRs (i.e, 149 SNRs) in the coverage of the Milky Way Imaging Scroll Painting (MWISP) CO survey and found that as many as 80 per cent of SNRs may be associated with molecular clouds. However, these surveys or studies are low resolution and/or low sensitivity.

Fortunately, there are recent large-scale surveys with higher resolution and better sensitivity, and recovery of extended emission such as the SARAO MeerKAT 1.3\,GHz Galactic Plane Survey \citep[SMGPS;][]{2024MNRAS.531..649G} that has imaged the Galactic Plane at radio wavelengths. It provides a well-selected and uniform sample of \ion{H}{ii} regions and SNRs from which a statistical study can be carried out. \cite{2025A&A...Bordiu} present the SMGPS extended source catalogue, containing 3326 and 263 of such extended sources of \ion{H}{ii} regions and SNRs, respectively. There are also the Structure, Excitation, and Dynamics of the Inner Galactic InterStellar Medium \citep[SEDIGISM;][]{2021MNRAS.500.3064S} and the $^{13}$CO/C$^{18}$O(3--2) Heterodyne Inner Milky Way Plane Survey \citep[CHIMPS;][]{2016MNRAS.456.2885R, 2019A&A...632A..58R} surveys of molecular clouds that can image the Galactic Plane at excellent resolution, revealing thousands of molecular clouds in the Milky Way \citep[e.g.][]{2021MNRAS.500.3027D, 2022MNRAS...Rani}. We focus our study on the SMGPS and SEDIGISM surveys for a number of reasons. First, the two large surveys of radio and molecular gas can provide a large enough sample of clouds that are being shocked or associated with \ion{H}{ii} regions and SNRs. With large samples we are able to explore intrinsic difference in the different populations of clouds associated with \ion{H}{ii} regions or SNRs and contrast them to unassociated clouds. Secondly, the SMGPS and SEDIGISM surveys can also trace the boundaries of the \ion{H}{ii} regions and SNRs with sufficient detail to be able to clearly identify molecular clouds that are interacting with the \ion{H}{ii} regions and SNRs. Moreover, as it is not possible from continuum imaging alone to determine the distance to \ion{H}{ii} regions and SNRs, identifying molecular clouds that are morphologically associated with individual SNRs allows their distance to be determined through kinematic distance methods. Finally, the SEDIGISM survey possesses the required spectroscopic imaging of the molecular gas to distinguish compressed regions of the gas through their broadened line profiles. This is significant as it is key in the estimation of the clouds' physical properties such as mass and size. It has been demonstrated that the SEDIGISM clouds have clumps within them with masses ranging from 10 M$_\odot$ to $10^5$ M$_\odot$ \citep[e.g.][]{2009A&A...504..415S, 2018MNRAS.473.1059U}.This could be another factor to consider as the star formation in some of these clumps may have been triggered by interaction with nearby \ion{H}{ii} regions or SNRs.

In this pilot study, we present a large scale study of triggered star formation potential as a result of the associations of \ion{H}{ii} regions and SNRs with molecular clouds from both the SMGPS and the SEDIGISM surveys, respectively, to better characterise the impact upon molecular clouds. Similarly, we examine the impact of massive stars on molecular clouds by combining the extensive radio survey with a compatible survey of molecular clouds. In Section \ref{sect: Data}, we give a brief description of both the SEDIGISM and the SMGPS surveys. In Section \ref{subsec: Analysis}, we detail the data analysis and methods employed to identify the best matches: associated SEDIGISM clouds and the interacting SMGPS extended \ion{H}{ii} regions and SNRs. We present our results in Section \ref{sec: Results}, with particular emphasis on the statistics of associated and unassociated SEDIGISM molecular clouds (Section \ref{subsec:Statistics of shocked vs unshocked clouds}), focusing on the molecular clouds' physical properties and the scaling relationships. A general discussion is provided in Section \ref{sect: Discussion}, and finally, we present our conclusions in Section \ref{sec: Conclusion}.

\section{Data}
\label{sect: Data}
We use data from the SARAO MeerKAT 1.3\,GHz Galactic Plane Survey (SMGPS)\footnote{\url{https://doi.org/10.48479/3wfd-e270}}, which is fully described in \cite{2024MNRAS.531..649G}. The SMGPS observed the Galactic Plane at 1.3\,GHz with an angular resolution of 8 arcseconds and a root-mean-square (RMS) sensitivity of $\sim$10--20 $\mu$Jy/beam. The observations were conducted between July 2018 and March 2020 from the 64-antenna MeerKAT array in the Northern Cape Province of South Africa \citep{Jonas:2018Jr,2020ApJ...888...61M,2024MNRAS.532.2842K} using the L-band (856--1712 MHz) receiver system with 4096 channels. The survey covers a wide portion of the first, second, and fourth Galactic quadrants ($\ell=2^{\circ}$--$61^{\circ}$, $251^{\circ}$--$358^{\circ}$, $|b|\leq 1.5^{\circ}$). Our pilot study focuses on the moment 0 maps of the G342.5 tile within the fourth Galactic quadrant of the survey with a Galactic longitude range of $341^{\circ}\leq \ell \leq 344^{\circ}$ and an approximate Galactic latitude of $|b|\leq 1.5^{\circ}$. It also uses the catalogue of SMGPS extended sources\footnote{\url{https://doi.org/10.48479/t1ya-na33}},%
\footnote{\url{http://cdsweb.u-strasbg.fr/cgi-bin/qcat?J/A+A/}},%
\footnote{\url{https://cdsarc.cds.unistra.fr/viz-bin/cat/J/A+A/695/A144}}
\citep{2025A&A...Bordiu}. The catalogue reveals the detailed structure of hundreds of \ion{H}{ii} regions and SNRs and comprises of 16538 extended and diffuse radio sources, of which about 24 per cent of them are directly connected to known Galactic objects: 3326 sources associated with \ion{H}{ii} regions, 263 with SNRs, 215 with planetary nebulae (PNe), 20 with luminous blue variables (LBVs), 7 with Wolf-Rayet (WR) stars and 59 objects with multiple associations; the rest correspond to candidate extragalactic sources (33 per cent) or unclassified objects (43 per cent). The G342.5 tile used in this pilot study contains 537 extended sources in total. Of these, 98 are classified as \ion{H}{ii} regions, 5 as SNRs, 7 as planetary nebulae (PNe), 213 as extragalactic galaxies, and 214 remain unclassified. For our analysis, we focus on the \ion{H}{ii} region and SNR populations within this field.

\begin{figure*}
    \centering
    \includegraphics[width=\textwidth, height=\linewidth]{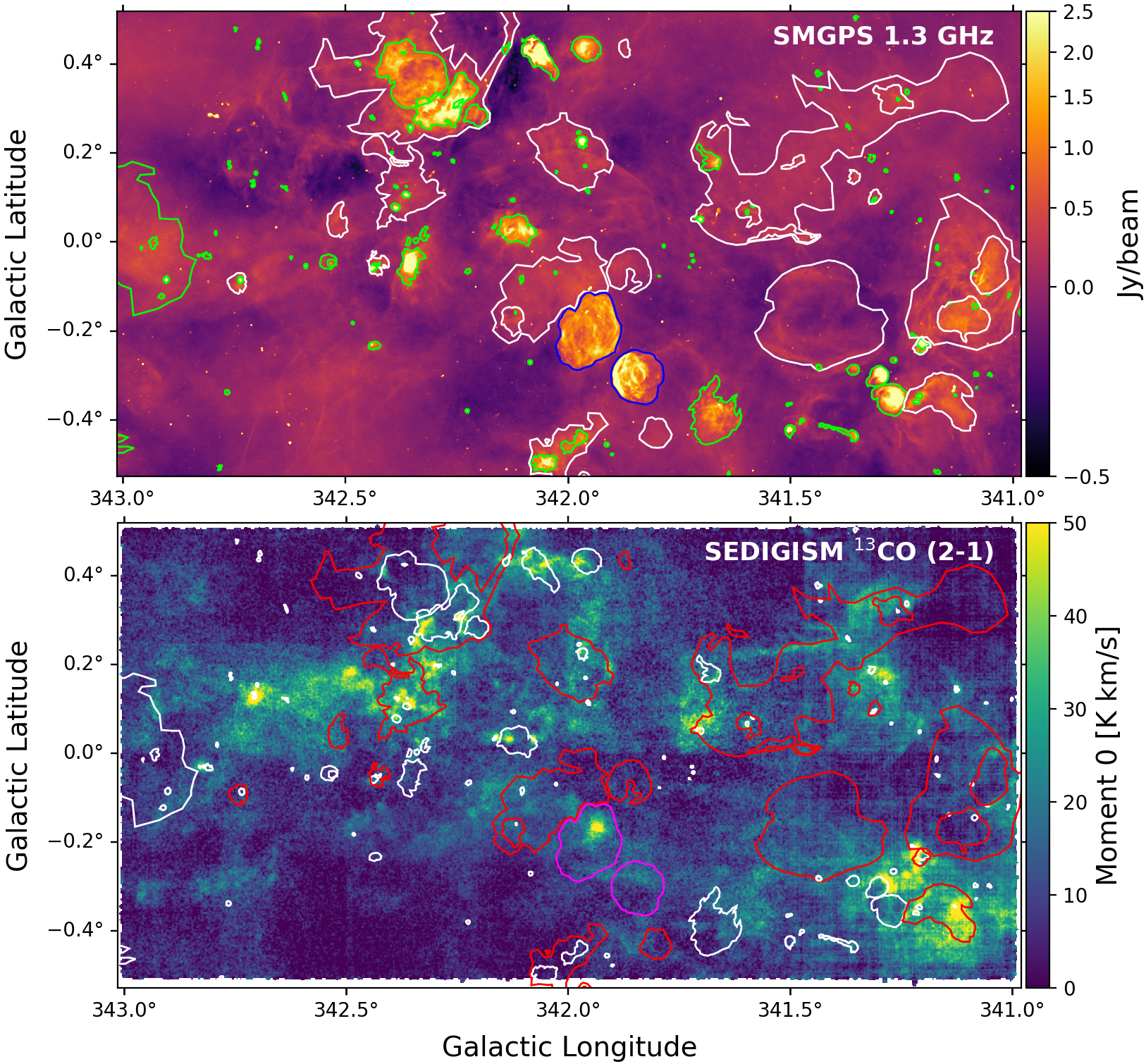}
    \caption{Top: A section of SARAO MeerKAT 1.3\,GHz Galactic Plane Survey intensity image within Galactic longitude range of approximately 341.0 to 343.0 degrees and Galactic latitude $ |b|\leq 0.5$ degrees, displayed on a square-root intensity scale. The green, white, and blue polygons outline SMGPS extended \ion{H}{ii} regions, diffuse \ion{H}{ii} region, and supernova remnants (SNRs), respectively. Bottom: Zeroth moment map of the G342 SEDIGISM $^{13}$CO (2--1) tile, integrated between\num{-150} and \num{50}\,\kms, on a linear intensity scale. The same extended SMGPS sources are overlaid on it with white, red, and magenta polygons identifying extended \ion{H}{ii} regions, diffuse \ion{H}{ii} regions and SNRs, respectively.}
    \label{fig:SMGPS_SEDIGISM}
\end{figure*}

The SEDIGISM survey was conducted during 2013–2017 \citep{2017A&A...601A.124S, 2021MNRAS.500.3027D, 2021MNRAS.500.3050U,2021MNRAS.500.3064S} and was observed with the SHFI single-pixel instrument of the 12m Atacama Pathfinder Experiment \citep[APEX;][]{2006A&A...454L..13G,2009A&A...504..415S,2022A&A...658A.160Y}. The survey imaged the Galactic Plane at approximately 30-arcsecond resolution, revealing more than 10,000 molecular clouds in the Milky Way. Data cubes of $^{13}$CO (2--1) surrounding each of these \ion{H}{ii} regions and SNRs were obtained from the public SEDIGISM web page\footnote{\url{https://sedigism.mpifr-bonn.mpg.de/index.html}}. Composite images of the molecular gas and radio emission were constructed to identify morphological (i.e. shape) associations between the \ion{H}{ii} regions and SNRs. We use a SEDIGISM merged catalogue \citep{2021MNRAS.500.3027D} of 10663 molecular clouds, which are hereafter referred to as SEDIGISM molecular clouds \citep[][]{2022A&A...663A..56N, 2022A&A...664A..84N}. We also make use of the cloud assignment masks from the {\sc scimes} \citep{2015MNRAS.454.2067C} extraction of molecular clouds within SEDIGISM by \cite{2021MNRAS.500.3027D}, which uses clustering techniques to group dense regions of gas, focusing on finding connected structures in spatial and velocity dimensions with similar emission properties. This work is based on the SEDIGISM tile centred at G342, that covers the region between $340.98^{\circ} \leq \ell \leq 343.01^{\circ}$ and the Galactic latitude of $|b|\leq 0.5^{\circ}$, containing 268 molecular clouds, and has 0.25\,\kms--wide velocity channels between -200 and 200\,\kms.

\section{Cross-matching SMGPS extended sources and SEDIGISM molecular clouds}
\label{subsec: Analysis}

\begin{figure*}
    \centering
    \includegraphics[width=\textwidth]{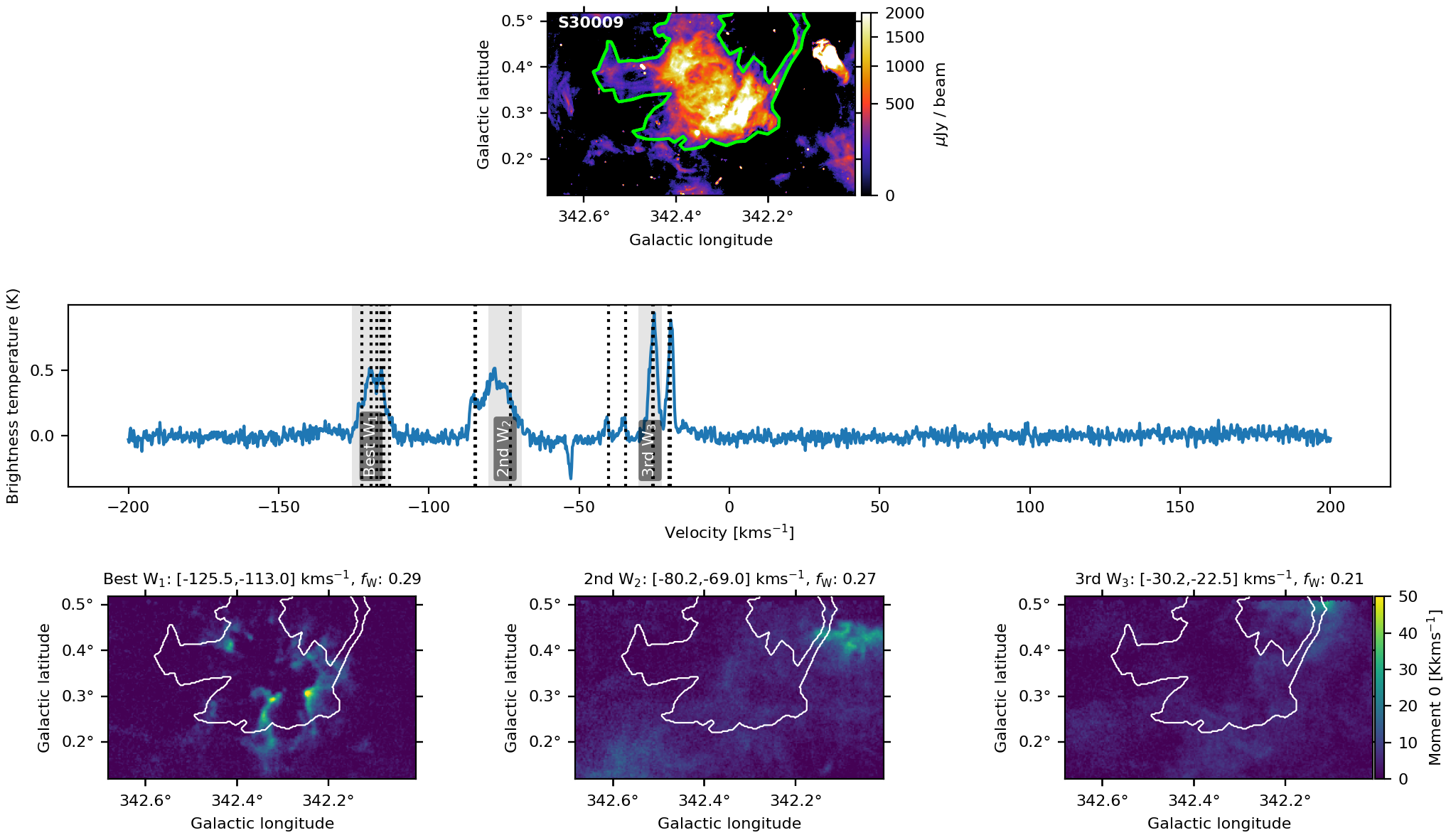}
    \caption{An example of the SMGPS-SEDIGISM association methodology for the extended SMGPS radio source S30009 \citep{2025A&A...Bordiu}.
    Top panel: SMGPS intensity map with the lime outline tracing the region of the source on a square-root intensity scale. Middle panel: Mean $^{13}$CO (2--1) spectrum extracted over the entire source area, showing the velocity-coherent emission windows. The best, second best, and third best-matching windows, ranked by the emission fraction (\fw), are represented as Best $\mathrm{W_1}$, 2nd $\mathrm{W_2}$, and 3rd $\mathrm{W_3}$, respectively. Vertical dashed lines in black colour indicate the velocity components of SEDIGISM molecular clouds intersecting the source region. Bottom panels: Moment-0 maps of the three most significant emission windows. The SMGPS source mask is shown as a white polygon, and the spectral windows' velocity ranges and emission fractions are given above each panel.}
    \label{fig:S30009_average_spectrum}
\end{figure*}

In this study we wish to identify physical associations between SMGPS extended 1.3\,GHz continuum sources from the SMGPS with molecular clouds traced by the spectral survey SEDIGISM. Our vantage point within the disc of the Milky Way presents difficulties for making associations between features identified within continuum and spectral data, because any given sight-line close to the midplane of the inner Galaxy (i.e. within the Solar circle) has a high probability of producing chance alignments of sources located within different spiral arms segments at different distances, and with different or overlapping radial velocities. For some fraction of the SMGPS, the extended sources are \ion{H}{ii} regions that also appear in the Wide-field Infrared Survey Explorer (WISE) Catalogue\footnote{\url{http://astro.phys.wvu.edu/wise}} of Galactic \ion{H}{ii} Regions \citep[][]{2014ApJS..Anderson(2), 2015ApJS..Anderson(4), 2018ApJS..Anderson(3)}, and which have direct radial velocity measurements from observations of radio recombination lines (RRLs). However, this represents only a minority of the SMGPS sources, and we therefore developed a method to assign the most appropriate velocity to each SMGPS source by cross-matching to molecular emission from SEDIGISM $^{13}$CO (2--1) data. Importantly, \ion{H}{ii} regions and SNRs are not necessarily expected to share their morphology with the surrounding molecular gas: in some cases the ionized plasma may expand and drive out gas from its original location, creating structures that are somewhat anti-correlated, though related. Our method is therefore devised to allow robust and appropriate velocity assignment while accounting for cases where multiple groups of molecular clouds lie along the same line of sight. Hereafter we refer to all associations between SMGPS sources and SEDIGISM molecular clouds as \emph{complexes} because by definition the association includes, at minimum, an extended radio source and a molecular cloud. These complexes may be physically interacting systems, though we caution that our method cannot fully exclude chance line-of-sight coincidences.

We began by identifying all members of the SMGPS extended source catalogue \citep{2025A&A...Bordiu} that fall within the G342.5 SMGPS tile. A total of 537 extended sources were initially identified within this tile, which is larger in area than the G342 SEDIGISM tile. Once we restricted this sample to the area overlapping the SEDIGISM tile, the number of sources was reduced to 191 (2 SNRs, \ion{H}{ii} regions (64), unclassified sources (93) and others are classified as galaxies (31) and PNe (1) which have not been included in this study) which we illustrate in Figure \ref{fig:SMGPS_SEDIGISM}. These 191 SMGPS sources were subsequently used in our attempt to associate them with all molecular clouds identified in the corresponding SEDIGISM tile. Each SMGPS outline was used to construct a 2D source mask on the SEDIGISM pixel grid. This mask was then applied to extract the average $^{13}$CO(2--1) spectrum across the entire area of the source, rather than just a single central position. This produces a representative spectrum that captures the kinematics of the molecular gas along the line-of-sight (see Figure \ref{fig:S30009_average_spectrum}). \footnote{We also tested spectra averaged over the rim pixels of the masks, but found the approach to be too strongly biased toward the low-intensity, poorly defined edges of sources.}

The same binary mask was used to extract the corresponding section of the {\sc scimes} assignment cube \citep[][]{2015MNRAS.454.2067C, 2019MNRAS.483.4291C, 2021MNRAS.500.3027D}. This allowed us to identify velocity-coherent emission regions, that we call velocity windows (see Figure \ref{fig:S30009_average_spectrum}), within the extracted spectrum (that contains the same voxels as the extracted cube) and to link them to catalogued SEDIGISM molecular clouds. The velocity windows were identified as velocity ranges with consecutive channels that contain any non-zero value in the assignment cube. Such windows may contain a single molecular cloud, or multiple clouds that overlap in velocity. For each velocity window in the spectrum, we determined its relative significance by calculating the fraction of the total integrated intensity within that window, which we called the  window \emph{emission fraction} (\fw): 
 
\begin{equation}
\hspace{10em}
    f_{\mathrm{W}_i} = \frac{W_{i}}{\sum_{i=1}^{N} W_{i}},
    \label{eq:confidence}
\end{equation}

\noindent where \( W_i \) is the integrated intensity of the \( i^{th} \) emission window, and \(N\) is the total number of emission windows identified in the extracted spectrum.

The emission fraction is a proxy for the fraction of the total column density in the different velocity windows, and we are assuming that the most likely physical association is between the SMGPS source and the highest column density molecular gas. Therefore, in cases where we have multiple candidate velocity windows, we then rank them and select the window with the greatest value of \fw\ as the most likely the association. All SEDIGISM clouds falling within this best-matching emission window's velocity range are considered to be part of the association and therefore of the complex. The SMGPS source is then assigned the central velocity of the range (see Table \ref{tab: best_matches}).

\begin{figure}
    \centering
    \includegraphics[width=\linewidth]{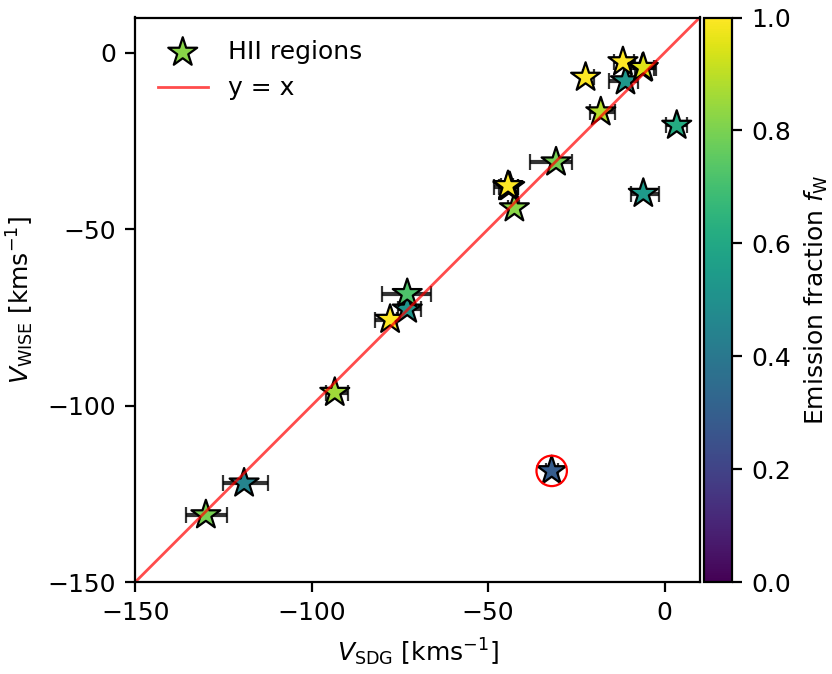}
    \caption{Scatter plot of the centroid velocities of matched \ion{H}{ii} regions from the SMGPS extended sources versus the systemic velocities of \ion{H}{ii} regions from the WISE catalogue of Galactic \ion{H}{ii} regions. The colour bar shows the emission fraction \fw\ of the associations between the interacting SMGPS extended sources with the associated molecular clouds from the best spectral emission windows. The red line indicates the one-to-one relation. The highlighted data point with red circle is a source with \fw$<0.3$, and represents less reliable association.}
    \label{fig:SMGPS_WISE}
\end{figure}

To test the validity of this indirect velocity-assignment method, we independently cross-checked the automated association process by comparing the CO-based velocities with the RRL velocities that we have access to. We compared the centroid velocities of clouds associated with \ion{H}{ii} regions to the systemic velocities of \ion{H}{ii} regions from WISE Catalogue of Galactic \ion{H}{ii} Regions \citep[][]{2014ApJS..Anderson(2), 2015ApJS..Anderson(4), 2018ApJS..Anderson(3)}. For this comparison, we selected only those WISE \ion{H}{ii} regions that lie within the G342.5 field and have a unique and single, unambiguous RRL velocity measurement (i.e, we reduced the WISE catalogue to only sources with a single RRL velocity). Figure \ref{fig:SMGPS_WISE} shows the comparison between our CO-based velocities ($V_\mathrm{SDG}$) and the RRL velocities. The data points cluster closely around the one-to-one line, demonstrating a strong correspondence between the two velocity measurements. The sample includes 20 WISE \ion{H}{ii} regions with RRL 
velocities ranging from -131 to -2.6 \kms, and the median absolute velocity discrepancy between the CO-- and RRL--based measurements is 2.96 \kms, indicating good agreement within the expected uncertainties. We highlight source(s) with \fw$<0.3$ using red circles; these represent less reliable matches where the dominant velocity window does not account for a sufficient fraction of the total emission.

\begin{figure}
    \centering
    \includegraphics[width=\linewidth]{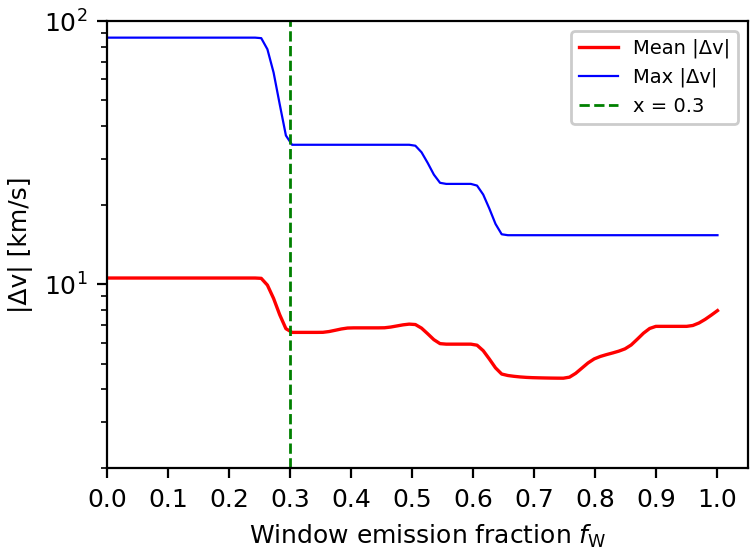}
    \caption{Mean (red) and maximum (blue) absolute velocity discrepancies (|$\Delta_v|$) between WISE RRL velocities and SMGPS-SEDIGISM associations, as a function of the emission fraction (\fw). The dashed green vertical line marks \fw$=0.3$, the threshold adopted in this study. Above this threshold, the mean velocity difference decreases from $\sim 10.5$ \kms\ to $\sim 6$ \kms, while the maximum difference drops from $\sim 86$ \kms\ to $\sim 35$ \kms. The decreasing trend suggests that a higher emission fraction corresponds to superior kinematic agreement, signifying a more reliable physical association.}
    \label{fig:mean_discrepancy_n_confidence}
\end{figure}

While Figure \ref{fig:SMGPS_WISE} demonstrates the validity of our velocity-assignment method, we further refined our working sample by examining how the velocity discrepancies depend on the emission fraction. Figure \ref{fig:mean_discrepancy_n_confidence} shows a plot of the mean absolute velocity difference (|$\Delta v$|, red curve) and the maximum |$\Delta v$| (blue curve) against the window's emission fraction, \fw. At \fw$\approx0.3$, the maximum discrepancy drops sharply from $\sim 86$ \kms\ to $\sim 35$ \kms, while the mean discrepancy decreases from $\sim 10.5$ \kms\ to $\sim 6$ \kms. Typical linewidths of RRLs in \ion{H}{ii} regions are 20--25 \kms\ \citep{2011ApJS..Anderson}. For \fw$\geq0.3$, our assigned velocities agree with RRL velocities well within this range on average, even though some mismatches of up to $\sim 35$\,\kms\ remain possible. This threshold therefore provides a good balance between velocity accuracy and maintaining a statistically useful sample size. In all subsequent analysis, we restrict our associations to complexes with  \fw$\geq0.3$ (dashed green line in Figure \ref{fig:mean_discrepancy_n_confidence}).

Overall, 159 of the 268 (60 per cent) SEDIGISM molecular clouds in the field intersect 131 SMGPS sources along the line of sight. Of these, 90 clouds (34 per cent) fall within the best-matching velocity windows and are therefore associated into complexes with the 131 extended SMGPS sources. Although around 90 per cent of the complexes contain only a single SEDIGISM cloud, their spectra almost always display multiple possible velocity components. This is illustrated in Appendix \ref{sec:complex_matches} (Figure \ref{fig:weighted_std_confidence}). Nevertheless, we found that this method generally produces morphologically sensible matches, as shown in the example of source S30009 (Figure \ref{fig:S30009_average_spectrum}). For this source the $^{13}$CO (2--1) emission traces dense pillars and a bubble rim of a size and shape that are clearly the most appropriate matches for the SMGPS emission compared to the alternatives.

\begin{table*}
\centering
\caption{Details of associations between SMGPS extended sources and the SEDIGISM molecular cloud counterparts that fall within the best emission windows. Columns 1--3 give the SMGPS source name,  centroid longitude and centroid latitude. Columns 4--7 give the name, centroid longitude, centroid latitude, and centroid velocity for the associated SEDIGISM molecular cloud. Column 8 gives the velocity range for the best emission or spectral window where all the associated clouds fall, column 9 gives the pixel coverage to quantify how much of a given SMGPS source is covered by SEDIGISM cloud(s), and column 10 gives an emission fraction (\fw) for the best matching velocity window.}
\resizebox{\hsize}{!}{
\begin{tabular}{cccccccccccccccc}
\hline
\hline
SMGPS & $\ell_{s}$ & $b_{s}$ & SEDIGISM & $\ell_{c}$ & 
$b_{c}$ & $v_\text{lsr}$ & Window's $v_\text{lsr}$ range & Pixel coverage & Emission fraction\\
(Source Name) & (deg) & (deg) & (Cloud Name) & (deg) & 
(deg) & (\kms) & (\kms) & - & ($f_W$)\\
(1) & (2) & (3) & (4) & (5) & (6) & (7) & (8) & (9) & (10) \\
\hline
S10002 & 341.37016 & 0.25870 & SDG341.396+0.2606 & 341.39584 & 0.26056 & -36.91 & -38.50, -35.25 & 0.111 & 0.480 \\
S10138 & 342.20381 & 0.67177 & SDG342.270+0.2943 & 342.26989 & 0.29431 & -119.15 & -122.50, -117.00 & 0.477 & 0.471 \\
S10191 & 341.31258 & 0.09381 & SDG341.307+0.1956 & 341.30691 & 0.19563 & -23.39 & -25.50, -22.00 & 1.000 & 0.539 \\
S10269 & 341.28561 & 0.15991 & SDG341.307+0.1956 & 341.30691 & 0.19563 & -23.39 & -26.00, -21.50 & 1.000 & 0.946 \\
S10313 & 341.27245 & 0.06736 & SDG341.293+0.0650 & 341.29314 & 0.06496 & -69.02 & -71.75, -67.25 & 1.000 & 0.636 \\
S10359 & 341.25796 & 0.32030 & SDG341.318+0.3339 & 341.31751 & 0.33387 & -77.76 & -79.25, -78.50 & 0.333 & 1.000 \\
S10419 & 341.23781 & 0.33703 & SDG341.318+0.3339 & 341.31751 & 0.33387 & -77.76 & -82.00, -75.75 & 1.000 & 1.000 \\
S10458 & 341.22593 & -0.20822 & SDG341.259-0.2767 & 341.25888 & -0.27674 & -44.35 & -46.25, -41.50 & 1.000 & 1.000 \\
S10474 & 341.21679 & -0.35932 & SDG341.215-0.3487 & 341.21457 & -0.34866 & -30.48 & -35.25, -28.75 & 1.000 & 0.726 \\
S10487 & 341.21804 & -0.21317 & SDG341.259-0.2767 & 341.25888 & -0.27674 & -44.35 & -46.25, -40.25 & 1.000 & 1.000 \\
S10498 & 341.20308 & -0.22909 & SDG341.259-0.2767 & 341.25888 & -0.27674 & -44.35 & -48.50, -40.00 & 0.965 & 1.000 \\
S10510 & 341.20880 & 0.04907 & SDG341.246+0.0301 & 341.24590 & 0.03011 & -76.25 & -78.25, -75.00 & 1.000 & 0.535 \\
S10707 & 341.14277 & -0.13745 & SDG341.209-0.1090 & 341.20894 & -0.10898 & -42.31 & -43.50, -41.75 & 1.000 & 1.000 \\
S10748 & 341.12961 & -0.34441 & SDG341.123-0.3523 & 341.12320 & -0.35231 & -41.66 & -43.75, -38.00 & 1.000 & 1.000 \\
S10753 & 341.12612 & 0.14439 & SDG341.101+0.1484 & 341.10128 & 0.14843 & -120.95 & -122.50, -117.50 & 1.000 & 0.590 \\
S10851 & 341.09079 & -0.06974 & SDG341.102-0.0793 & 341.10155 & -0.07925 & -46.98 & -48.25, -45.75 & 1.000 & 1.000 \\
S10870 & 341.08300 & -0.29729 & SDG341.123-0.3523 & 341.12320 & -0.35231 & -41.66 & -43.00, -41.50 & 0.111 & 1.000 \\
S10881 & 341.08161 & -0.32761 & SDG341.041-0.3593 & 341.04113 & -0.35927 & -32.23 & -32.50, -31.75 & 1.000 & 1.000 \\
S10936 & 341.05902 & 0.11352 & SDG341.101+0.1484 & 341.10128 & 0.14843 & -120.95 & -122.50, -122.25 & 0.000 & 1.000 \\
S10940 & 341.05608 & -0.11838 & SDG341.016-0.1252 & 341.01593 & -0.12517 & -42.55 & -46.25, -41.75 & 1.000 & 1.000 \\
S11073 & 341.00899 & -0.11997 & SDG341.016-0.1252 & 341.01593 & -0.12517 & -42.55 & -45.25, -41.00 & 1.000 & 1.000 \\
S11150 & 340.97854 & -0.15908 & SDG341.016-0.1252 & 341.01593 & -0.12517 & -42.55 & -43.00, -41.50 & 0.200 & 0.000 \\
S30007 & 342.43621 & 0.27675 & SDG342.458+0.2648 & 342.45803 & 0.26480 & -117.36 & -118.75, -118.50 & 0.000 & 1.000 \\
S30008 & 342.44140 & 0.28068 & SDG342.458+0.2648 & 342.45803 & 0.26480 & -117.36 & -119.00, -115.50 & 1.000 & 1.000 \\

$\vdots$ & $\vdots$ & $\vdots$ & $\vdots$ & $\vdots$ & $\vdots$ & $\vdots$ & $\vdots$ & $\vdots$ & $\vdots$\\
\hline\hline
\end{tabular}
}
\begin{minipage}{18cm}
\vspace{0.2cm}
\footnotesize \textbf{Notes:} Apart from the Pixel coverage and Emission fraction columns, the remaining quantities including the source names and their coordinates are taken from the SMGPS extended sources catalogue by \cite{2025A&A...Bordiu} and the cloud names, their coordinates, and centroid velocities are originally taken from SEDIGISM catalogue by \cite{2021MNRAS.500.3027D}. A full electronic version of this table will be available with this article.
\end{minipage}
\label{tab: best_matches}
\end{table*}

\section{Results}
\label{sec: Results}
In Section \ref{subsec: Associations} we present associations between the SMGPS sources (\ion{H}{ii} regions) and SEDIGISM molecular cloud
complexes. Individual regions are chosen to underline the complexity of some of the associations. In Section \ref{subsec:Statistics of shocked vs unshocked clouds} we compare the properties of associated SEDIGISM clouds to a control sample of clouds that are not associated or interacting with \ion{H}{ii} regions or SNRs. With a large and well selected sample of clouds associated and unassociated with \ion{H}{ii} regions and SNRs, we statistically investigate any differences in the physical and star-forming properties of the two samples.

\subsection{Associations between the SMGPS and molecular cloud complexes}
\label{subsec: Associations}

\begin{figure}
    \centering
    \includegraphics[width=\linewidth]{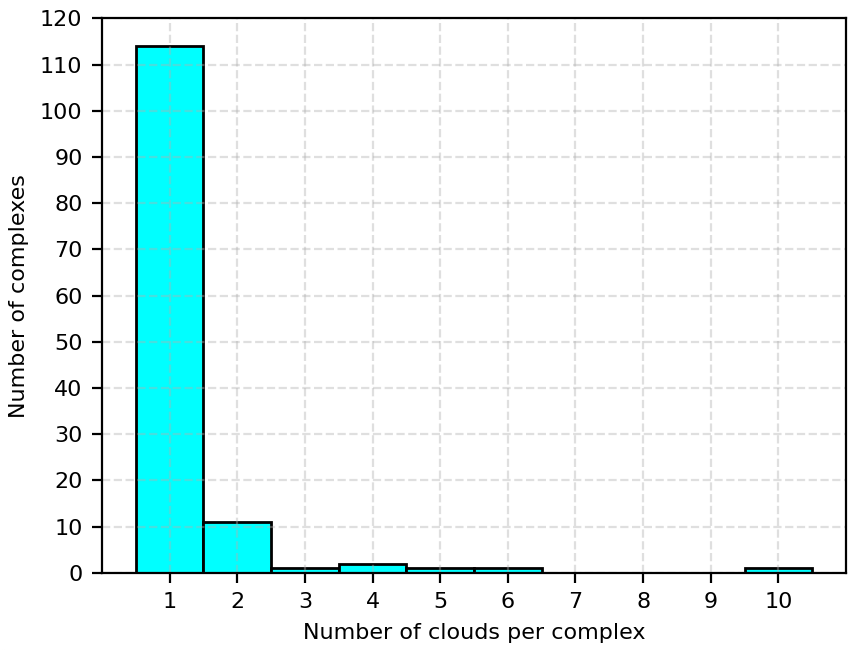}
    \caption{Histogram distribution of the number of SEDIGISM molecular clouds per complex, within the best-matching velocity window, associated with SMGPS sources.} 
    \label{fig:histogram of clouds}
\end{figure}

Figure \ref{fig:histogram of clouds} shows the distribution of the number of clouds associated with each SMGPS source. Out of 131 complexes, the majority (114; $\sim$87 per cent) are associated with a single molecular cloud. A smaller fraction (11 complexes; $\sim 8$ per cent) contains two clouds, while only 6 complexes in total (5 per cent) are associated with three or more clouds. The most extreme case involves one complex (e.g., S30026, see Table \ref{tab:clouds' physical properties}), which is linked to ten clouds. This distribution highlights that most SMGPS sources are linked to only one molecular cloud, while a minority are connected to larger, more complex environments.

\begin{figure}
    \centering
    \includegraphics[width=\linewidth]{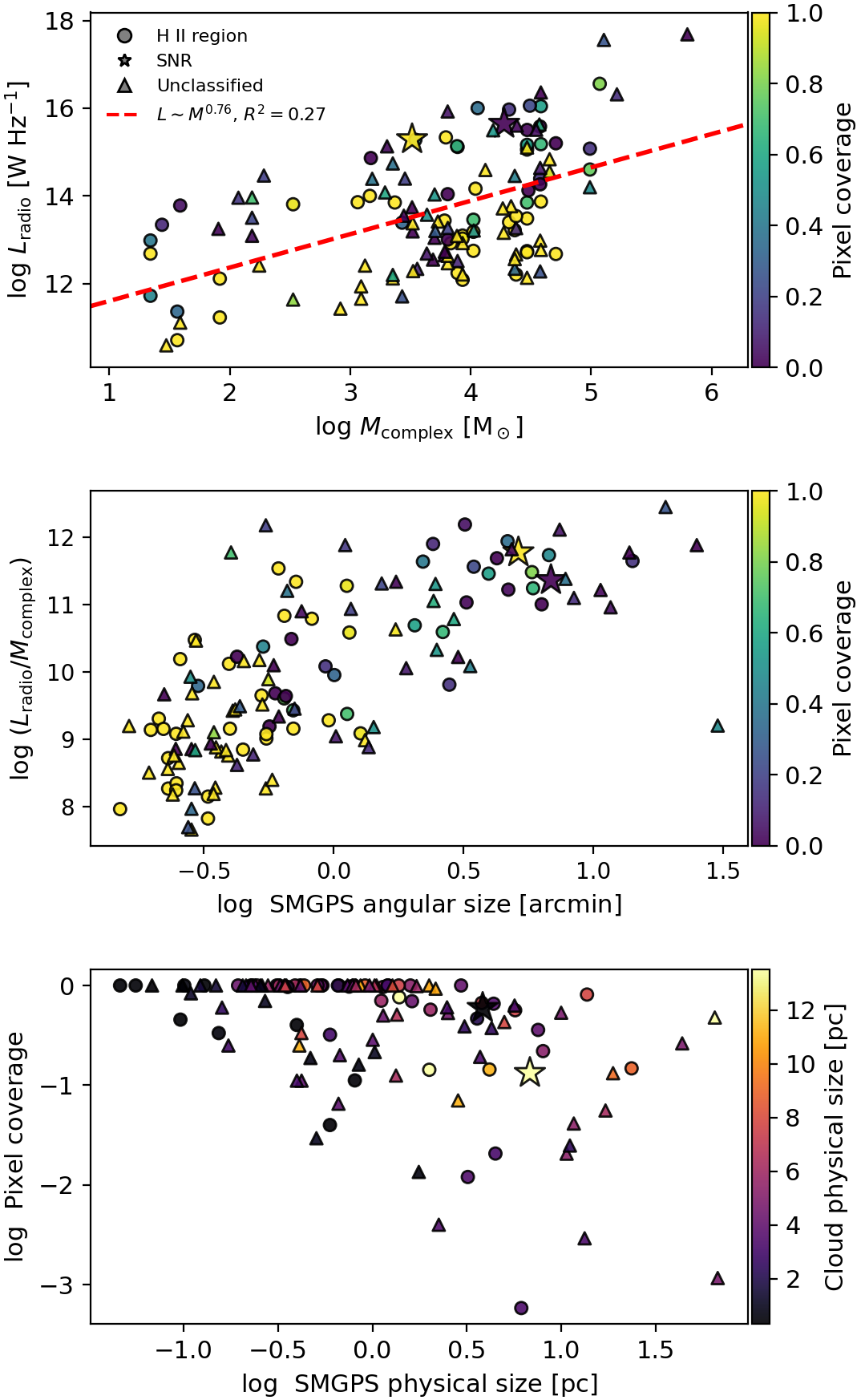}
    \caption{Top panel: Relationship between SMGPS radio luminosity (\lradio) and complex molecular mass (\mcomplex) for all associations. Sources are classified as \ion{H}{ii} regions (circles), supernova remnants (stars), and unclassified sources (triangles). The colour scale shows the cloud pixel coverage. The red dashed line indicates the best-fit linear regression, corresponding to a power-law relation $L_\mathrm{{radio}} \propto M_\mathrm{{complex}}^{0.76}$ with a coefficient of determination $R^{2}=0.27$.
    Middle panel: Relationship between ratio of radio luminosity to complex molecular mass (\lm) and SMGPS angular size, with points colour-coded by pixel coverage. Symbols are as in the top panel.
    Bottom panel: SMGPS source physical size (radius) as a function of pixel coverage of a SEDIGISM cloud on a SMGPS source, colour-coded by the physical size (radius) of the associated SEDIGISM cloud. Symbols are as in the top panel.}
    \label{fig:L_radio vs mass}
\end{figure}

The dominance of single-cloud associations likely reflects the compact nature of most SMGPS sources. These objects are small in angular size, with approximately 70 per cent of the \ion{H}{ii} regions having radii smaller than 2 arcmin (and nearly 80 per cent smaller than 3 arcmin; also see Figure \ref{fig:L_radio vs mass}, middle panel), and are therefore consistent with relatively young \ion{H}{ii} regions that remain confined to their natal molecular cloud \citep{2021MNRAS.500.3050U}, rather than tracing isolated environments per se. By contrast, the small population of multi-cloud complexes is significant: although they are rare, they contribute $\sim 85$ per cent of the integrated SMGPS flux, $\sim 95$ per cent of the radio luminosity, and $\sim 80$ per cent of the total angular area. Therefore, despite their rarity in number, these larger associations thus tend to dominate the emission budget of the region since they are the largest and brightest SMGPS sources, and are thus likely to be the most evolved, spatially extended \ion{H}{ii} regions containing the brightest massive clusters or interacting with multiple nearby molecular clouds. Such environments are natural laboratories for investigating feedback-driven cloud disruption/compression and the potential for triggered star formation. However, a complete picture of triggering requires examining both ends of this evolutionary sequence: the large multi-cloud complexes that dominate the emission, and the compact, single-cloud \ion{H}{ii} regions that may represent the earliest stages of feedback-driven star formation. 

The top panel of Figure \ref{fig:L_radio vs mass} shows the relationship between SMGPS radio luminosity (\lradio) and the total molecular cloud mass (\mcomplex) for each complex. The sample includes \ion{H}{ii} regions, supernova remnants, and unclassified sources. Pixel coverage (the fraction of a given SMGPS source covered by an associated SEDIGISM cloud in terms of pixels) is encoded by the colour scale. A positive correlation is  seen, described by a best-fit power-law relation $L\propto M^{0.76}$,  with a coefficient of determination $R^{2}=0.27$. Thus, more luminous sources are generally associated with more massive complexes, as expected if more massive clouds host more massive star formation. 

We also note a systematic behaviour with pixel coverage: for a given complex mass, lower pixel coverage tends to correspond to higher luminosities. This suggests an evolutionary effect -- younger, more embedded \ion{H}{ii} regions show high pixel coverage, while evolved regions have excavated molecular material as they expand, leading to lower coverage. The result is that the luminosity-to-mass ratio (\lm) increases with evolutionary stage, consistent with trends seen at the clump scale \citep[e.g.][Figure 24]{2018MNRAS.473.1059U}. The two supernova remnants lie on the high-luminosity edge of the relation, but with opposite pixel coverages; this may indicate that their associations are chance, and with only two objects no firm conclusions can be drawn.

From the apparent trend that we notice in the top panel, which may relate to source evolution, we therefore want to investigate this further by plotting \lm\ versus size.  In the middle panel of Figure \ref{fig:L_radio vs mass}, a correlation is shown such that we find that the complex luminosity–to–mass ratio (\lm) generally increases with the angular extent of the SMGPS sources. While there is appreciable scatter, more extended sources (which may generally be interpreted as more evolved \ion{H}{ii} regions) tend to show higher \lm\ values, suggesting that larger angular sizes are associated with enhanced radio emission relative to their cloud masses. This is consistent with the expectation that \ion{H}{ii} regions expand as they evolve. We also expect that multiple compact \ion{H}{ii} regions will merge. The bottom panel of Figure \ref{fig:L_radio vs mass} shows a negative correlation between pixel coverage and the physical size of SMGPS sources. Larger (and therefore more evolved) \ion{H}{ii} regions have reduced overlap with molecular clouds, which is what we expect and is consistent with the expansion eroding their natal environments.

Finally, the scatter present in all three panels (Figure \ref{fig:L_radio vs mass}) is partly physical, but it also reflects limitations of the cloud association process. Because the total mass of all intersecting SEDIGISM clouds in the best-matching velocity window is included, rather than only the mass contained within the SMGPS boundaries, some of the mass factored in these measurements may lie at significant offsets from the radio source. This can lead to cases where massive GMCs contribute to multiple radio–cloud associations, or where part of the cloud mass is unrelated to the current star-forming region. Such effects introduce a degree of random scatter into the measured relations. 

\begin{figure*}
    \centering
    \includegraphics[width=\textwidth]{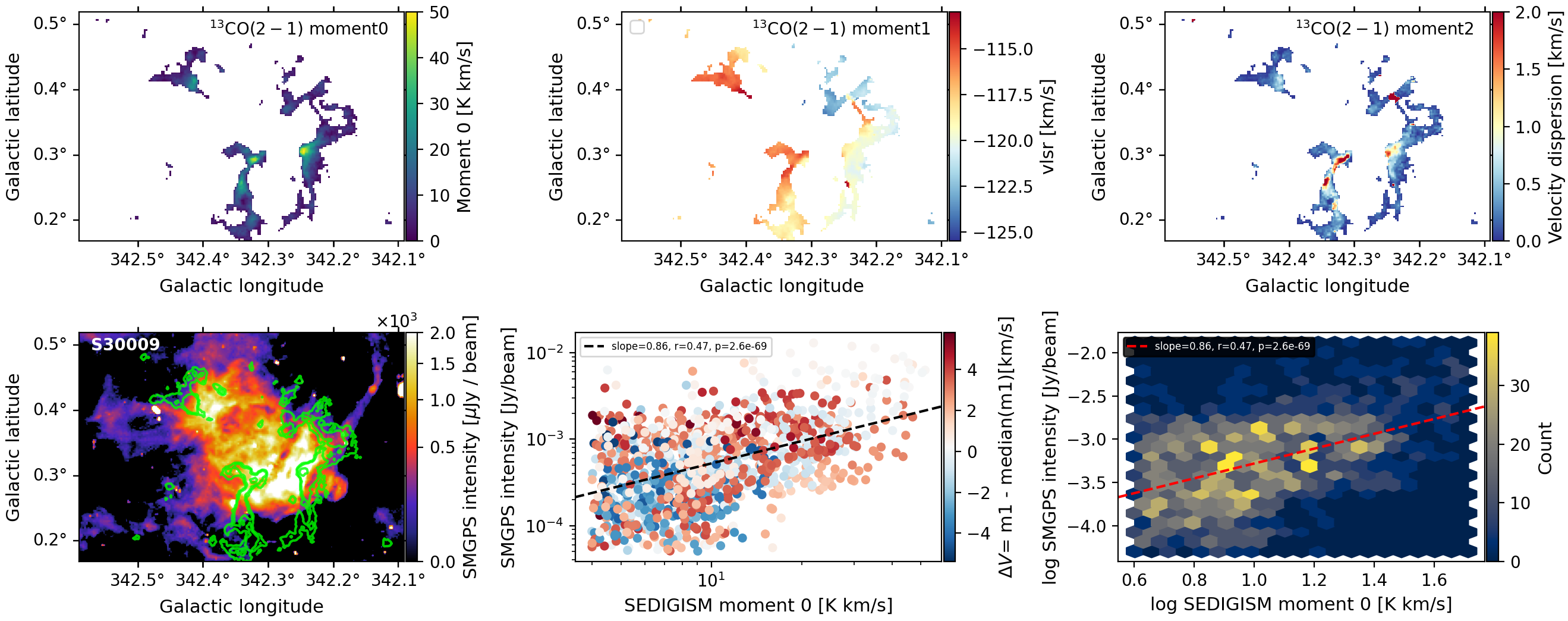}
    \caption{Comparison of SMGPS and SEDIGISM data for SMGPS source S30009. Top row: $^{13}$CO (2--1) moment maps integrated between -126 and -113 \kms. From left to right: zeroth moment (integrated intensity), first moment (intensity-weighted velocity), and second moment (velocity dispersion). Bottom row: SMGPS intensity image (left), overlaid with $^{13}$CO moment 0 contours; pixel-by-pixel scatter plot of SMGPS intensity versus $^{13}$CO moment 0 (middle), colour-coded by velocity offset from the median value, with gradient, Pearson correlation coefficient and $p$-value in the legend; hexagonal histogram (right) of SMGPS intensity versus $^{13}$CO moment 0 on a pixel-by-pixel basis.}
    \label{fig:complex_pixel-pixel_scatter}
\end{figure*}

\begin{figure*}
    \centering
    \includegraphics[width=\textwidth]{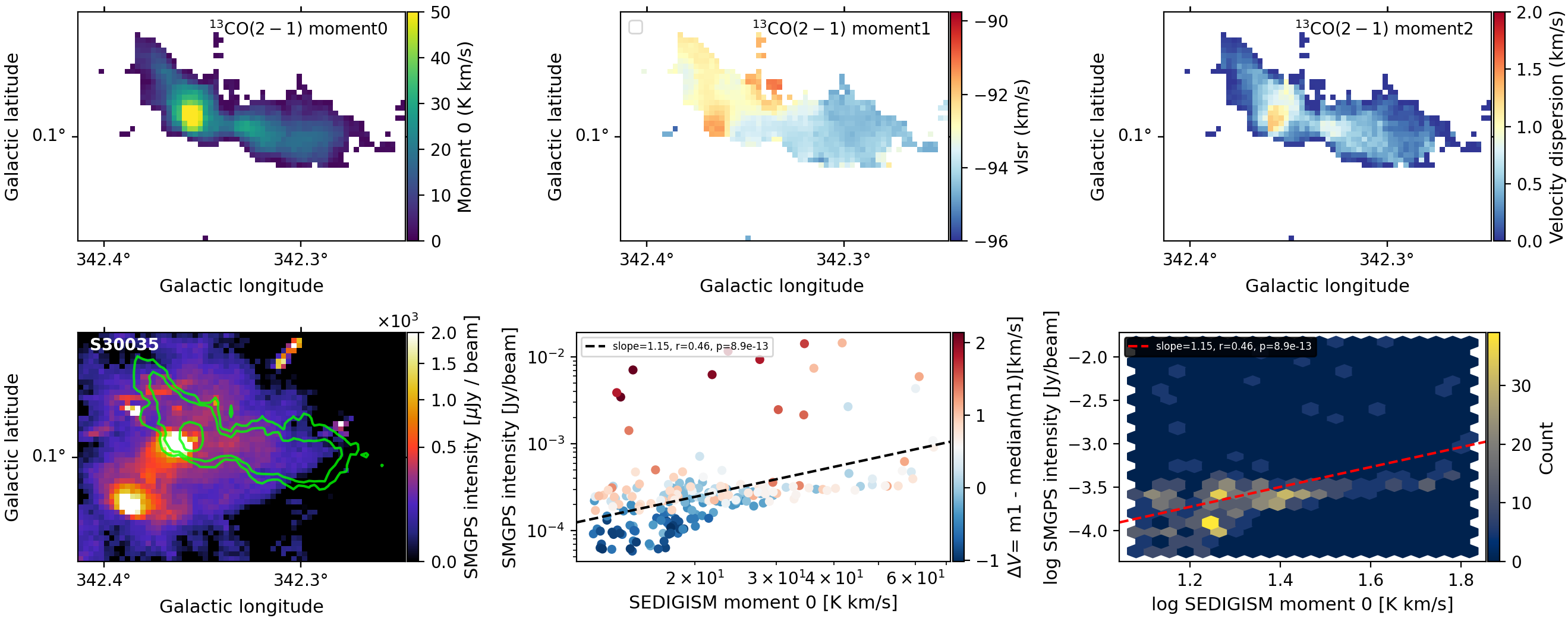}
    \caption{Same as Figure \ref{fig:complex_pixel-pixel_scatter} for the SMGPS source S30035, with moment maps integrated between -96 and -90\,\kms.}
    \label{fig:compact_pixel-pixel_scatter}
\end{figure*}

As an example, in Figure \ref{fig:complex_pixel-pixel_scatter} we show the zeroth, first, and second moment maps of all SEDIGISM molecular clouds associated with the SMGPS source S30009, integrated over the velocity range between -126 and -113\,\kms. The moment zero map (tracing H$_2$ column density) reveals the distribution of molecular gas, with the emission concentrated in filamentary structures across the \ion{H}{ii} region, S30009. The first moment map highlights a velocity gradient within the centroid velocity range, with gradients suggestive of internal motions or large-scale flows. The molecular cloud velocities are higher (-113\,\kms) near the central source and towards the northwest and decrease (to about -126\,\kms) outward towards the northeast. This is consistent with the expansion of an ionization front or shock wave driving the gas outward. The second moment map (velocity dispersion; top right) shows regions of high velocity dispersion, particularly near the boundaries of the clouds or particularly along dense ridges, consistent with regions of kinematic broadening. The bottom left panel indicates intensity-velocity gradient alignments, such that the continuum emission aligns with peaks of CO emission, strongly indicating an association between ionized gas and its parental molecular cloud. The region surrounding the source S30009 appears to have the alignment of SMGPS intensity gradients and velocity gradients of the clouds. This provides critical insights into the physical processes at play, most likely indication of turbulence or expansion along the line of sight due to the change in velocity dispersions or large linewidths as observed in the second moment map. We also have an instance of a large SMGPS intensity gradient running perpendicular to the ridge-like filamentary cloud that runs along its edge.

A more quantitative assessment of interaction is provided by the pixel-by-pixel comparison (Figure \ref{fig:complex_pixel-pixel_scatter}, bottom middle). A positive correlation between SMGPS intensity and $^{13}$CO (2--1) zeroth moment is evident, with a fitted slope of 0.86 and a Pearson correlation coefficient of $r=0.47$ ($p=2.6\times 10^{-69}$). This demonstrates that regions of higher molecular gas column density (traced by CO) are systematically associated with stronger 1.3\,GHz continuum emission, indicating that the ionized gas arises preferentially in the densest parts of the molecular cloud--where the massive stars have recently formed and begun ionizing their surroundings. The colour-coding by velocity offset shows that pixels with large velocity deviations are scattered around the trend but still follow the overall correlation, suggesting that the kinematic complexity does not erase the dust–gas coupling. The hexbin plot (bottom right) reinforces this trend statistically, highlighting that the bulk of the pixel population follows the same positive relation between gas and dust intensities.

In Figure \ref{fig:compact_pixel-pixel_scatter} we show another example, with the moment 0 map (top left)  showing the molecular gas distribution around the compact \ion{H}{ii} region, S30035, which is embedded within the centrally concentrated CO emission, revealing a clumpy and elongated morphology with bright integrated intensities up to $\sim 50$ K \kms. The moment 1 map (top middle), has a relatively smooth velocity field with systematic gradients, and the velocity dispersion map (moment 2, top right) highlights broadened regions (though lower levels of broadening compared to S30009) near the interface with the ionized gas, consistent with feedback-driven motions. The compact SMGPS \ion{H}{ii} region, S30035, exhibits a more centrally concentrated distribution of continuum intensity (bottom left). The continuum emission aligns with the peaks of CO emission, strongly indicating an association between the ionized gas and its parent molecular cloud.

The pixel-by-pixel scatter plot (Figure \ref{fig:compact_pixel-pixel_scatter}, bottom middle) between SMGPS intensity and SEDIGISM CO moment 0 reveals a statistically significant correlation. A best-fit slope of 1.15 suggests that the intensity of the radio continuum increases more than linearly with the integrated intensity of CO. The correlation coefficient of $r=0.46$ indicates a moderate but significant correlation ($p=8.9\times 10^{-13}$). The color-coding by velocity offset shows that pixels with higher continuum intensity are preferentially associated with regions closer to the systemic velocity, while pixels at large velocity offsets contribute less to the correlation. The hexbin map (bottom right) confirms that the majority of pixels cluster around the fitted trend, reinforcing the robustness of the relation despite scatter.

Both sources illustrate clear examples of molecular–ionized gas coupling, yet they appear to represent different evolutionary stages. In Figure \ref{fig:complex_pixel-pixel_scatter} (bottom-left panel),  the morphology shows fragmentation at the peripheries of the large, centralised \ion{H}{ii} region, S30009. It is suggested that this could be a sign of collect and collapse process, particularly at the centres and along borders of the sources. This process appears to be happening as the source (\ion{H}{ii} region) is completely surrounded by a ring of 6 molecular clouds (see also Figure \ref{fig:S30009_average_spectrum}), with the emissions arising from the gas and dust surrounding the \ion{H}{ii} region.  This interpretation aligns with previous studies that discuss the star formation by the collect and collapse process as a crucial driver of star formation at the boundaries of expanding \ion{H}{ii} regions \citep[e.g.][]{1998ASPC..148..150E, 2005A&A...433..565D, 2006A&A...446..171Z, 2007A&A...472..835Z, 2007MNRAS...Dale, 2008ASPC..Deharveng, 2010A&A...Zavagno, 2010ApJ...Peters, 2011A&A...Brand, 2017A&A...Figueira}. The $^{13}$CO emission maps further reveal dense fragments that are evenly spaced within the molecular ring surrounding the \ion{H}{ii} region and the surrounding gas appears as if being swept into dense shells that are becoming gravitationally unstable and may be collapsing to form a new generation of massive stars \citep{2009A&A...494..987P}. In contrast, the small SMGPS source, S30035 (Figure \ref{fig:compact_pixel-pixel_scatter}, bottom-left panel) appears more compact and deeply embedded in the dense parts of the host clumped molecular cloud (SDG342.331+0.1096). The cloud has 0.13, and 1.35 (L$_\odot$/M$_\odot$), DGF and SFE measurements, respectively. It is also massive (5080\,M$_\odot$), with a high surface density (202\,M$_\odot$\pc), a high velocity dispersion (1.18 \kms, which can also be observed within the central area of the moment 2 map), and a very low virial parameter ($\alpha_\text{vir}=0.28$); key physical factors supporting that the complex is in a relatively early stage of star formation \citep{2021MNRAS.500.3050U}.

\subsection{Statistics of associated and unassociated SEDIGISM molecular clouds}
\label{subsec:Statistics of shocked vs unshocked clouds}
In this section, we present a statistical analysis of the SEDIGISM molecular clouds associated with SMGPS \ion{H}{ii} regions and supernova remnants (SNRs), together with a control sample of the molecular clouds that are not associated with SMGPS sources, known hereafter as unassociated molecular clouds. There are a total of 268 clouds within the SEDIGISM tile. The clouds intersect 131 sources from the SMGPS tile of which there are only 2 SNRs and the rest are \ion{H}{ii} regions (57) and unclassified sources (72). We identify 90 of these clouds as being associated with 131 of the SMGPS extended sources and only 2/90 clouds are associated with the two SNRs (further discussed in Section \ref{subsec: SNRs}). The remaining 178 clouds form our unassociated sample and should therefore be unaffected by feedback. In Section~\ref{subsec: hist_distribution}, we compare the masses, average gas surface densities, linewidths (velocity dispersions), and virial parameters of the two categories of clouds to see if there are any statistical differences between the two populations. 

In Section~\ref{subsec:Scaling_relations}, we examine the scaling relationships of the clouds' physical properties such as size versus mass, size versus linewidth ($\sigma$) \citep[][]{1981MNRAS.194..809L, 1987ApJ...319..730S, 2001ApJ...562..348O, 2010ApJ...Kauffmann(a), 2010ApJ...Kauffmann(b), 2010ApJ...Kauffmann(c), 2013ApJ...779..185K, 2016ApJ...Nguyen-Luong, 2021MNRAS.500.3027D}, and average gas surface density ($\Sigma$) versus squared linewidth to size ratio ($\sigma^2/R$), known as the Heyer relation \citep{2009ApJ...699.1092H}. 

In Section~\ref{subsec:Radio_luminosity}, we investigate the correlations between the radio luminosity (\lradio) or physical size (radius) of the SMGPS sources and the associated molecular clouds' star formation efficiency (SFE), dense gas fraction (DGF), and gas surface density ($\Sigma$) for any possible indication of compression by the \ion{H}{ii} regions or SNRs. Here, we follow the definitions of \cite{2021MNRAS.500.3050U}, who matched ATLASGAL clumps to their parental SEDIGISM giant molecular clouds (GMCs) and derived both the DGF and the instantaneous star formation efficiency (SFE$_\mathrm{GMC}$). In this framework, SFE$_\mathrm{GMC}$ is given by the ratio of the total bolometric luminosity of ATLASGAL clumps within a GMC to the total mass of that GMC (\lclump/\mgmc). While $L$/$M$ for individual clumps is often used as an evolutionary stage indicator, \cite{2021MNRAS.500.3050U} explicitly define this cloud-scale \lclump/\mgmc\ as a measure of the instantaneous star formation efficiency. On the other hand, DGF$_\mathrm{GMC}$ is defined as the fraction of the GMC mass traced by compact dust emission, i.e. \mclump/\mgmc. We therefore use the values directly from \cite{2021MNRAS.500.3050U}, who provide SFE$_\mathrm{GMC}$ and DGF$_\mathrm{GMC}$ consistently for the SEDIGISM GMC sample, rather than re-estimating them in this work. In addition, we also look at the statistics of the two populations of clouds based on the presence or absence of a generic high-mass star formation (HMSF) indicator, which includes the presence of methanol masers \citep{2013MNRAS...Urquhart(a), 2015MNRAS....Urquhart}, \ion{H}{ii} regions \citep{2013MNRAS.435..Urquhart}, or massive young stellar objects \citep{2014MNRAS...Urquhart}. As such we adopt the classification of \cite{2021MNRAS.500.3027D}, who flagged SEDIGISM molecular clouds as having a HMSF tracer (=1) or not (=0), as per \cite{2014MNRAS...Urquhart}. Just as \cite{2021MNRAS.500.3027D} did not cross-match the SEDIGISM clouds with HMSF tracers directly, we also have not necessarily confined these tracers to the boundaries of SMGPS sources (or specify the precise evolutionary stage of the embedded sources or their spatial relationship to SMGPS emission); rather, they are used here as indicators of whether a molecular cloud is associated with one or more dense clumps that host massive mass star-forming activity. We combine this information with SFE (L/M), DGF values, and HMSF tracer to provide a statistical measure of a cloud’s star-forming potential.

\subsubsection{Distributions of cloud's mass ($M_\odot$), average gas surface density ($\Sigma$), linewidth ($\sigma_v$), and virial parameter ($\alpha_\text{vir}$)}
\label{subsec: hist_distribution}

\begin{figure*}
    \centering
    \includegraphics[width=\textwidth]{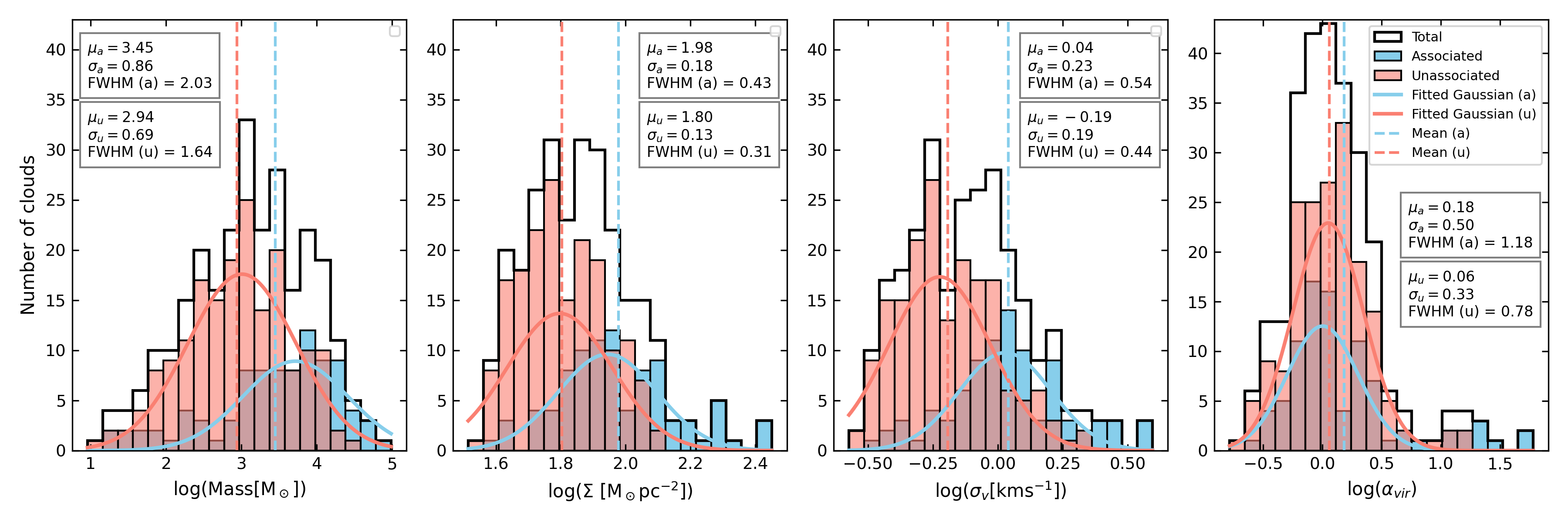}
    \caption{Histograms of four physical properties of both the associated and unassociated SEDIGISM molecular clouds. From left to right: Distribution of log of mass for the molecular clouds, distribution of log of deconvolved average gas surface density ($\Sigma$) of the clouds, distribution of log of clouds' velocity dispersion (linewidth) ($\sigma_v$), and distribution of log of deconvolved virial parameter($\alpha_\mathrm{vir}$) of the clouds, respectively. The Gaussian parameters are as shown and the bin size is 20 for each histogram. The values of mean mass, standard deviation, and Full Width at Half Maximum (FWHM) are as indicated with ${\textit{a}}$ and ${\textit{u}}$ representing associated and unassociated clouds, respectively. The third histogram in the back as a black outline, shows the whole (Total) sample (i.e. associated + unassociated samples).} 
    \label{fig:mass}
\end{figure*}

Figure \ref{fig:mass} presents histograms of four key physical properties—mass, average gas surface density, linewidth (velocity dispersion), and virial parameter—of both associated and unassociated SEDIGISM molecular clouds. The associated clouds exhibit a significantly larger mean mass of $\sim$9600 M$_\odot$ compared to the unassociated clouds'  $\sim$2500 M$_\odot$), as evidenced by the fitted Gaussian parameters. Besides having a larger mean mass, the associated clouds also have a wider distribution of masses, reflected in the Full Width at Half Maximums (FWHMs) of 2.03 and 1.64 dex for the associated and unassociated clouds, respectively, and the masses of the associated clouds extend to values an order of magnitude higher than the unassociated sample. Also of note is the fact that the lower-mass ends of the distributions are broadly similar. A Kolmogorov-Smirnov (K-S) test on the mass distribution yields a $p$-value of $1.5\times10^{-7}$, allowing us to reject the null hypothesis that both distributions are drawn from the parent sample and that the difference in mass distributions between the two categories of clouds is statistically significant.

Similarly, the distribution of the log-transformed average gas surface density reveals a higher mean for the associated clouds (104 M$_\odot$\pc) compared to the unassociated ones (67 M$_\odot$\pc), with a larger dispersion (0.43 dex versus 0.31 dex). This indicates that associated clouds tend to span a wider range of surface densities, including the highest values in the sample.
Generally, it implies that at higher surface densities the fraction of associated clouds increases, consistent with the expectation that massive star formation is more likely to occur in dense environments. However, there is no strict threshold, as associations are also found at lower surface densities—possibly because localised overdensities or compact clumps within the GMCs are not captured by the cloud-wide average surface density. The K-S test for surface density results in a $p$-value of $2.74\times10^{-11}$, which allows us to reject the null hypothesis and conclude that the difference between the two cloud populations is statistically significant.

Just like mass and average gas surface density, the distribution of the log-transformed linewidths (velocity dispersions, $\sigma_v$) shows that associated clouds have higher linewidths with a mean value of $\sim 1.26$ \kms\ compared to the unassociated clouds’ $\sim 0.71$ \kms. As shown by the fitted Gaussian parameters, the
associated population also exhibits a broader spread, with a FWHM
of 0.54 dex versus 0.44 dex for the unassociated population, suggesting that clouds associated with SMGPS sources not only tend to have larger internal turbulent motions, but also span a wider dynamical range of linewidths. The increased velocity dispersions are consistent with the picture of more active
environments, where stellar feedback from massive stars injects turbulence into the gas. A K-S test on the linewidth distributions confirms that the difference in velocity dispersion distributions between the two populations of clouds is statistically significant (with $p=7.85\times 10^{-14}$).

The virial parameter, $\alpha_\text{vir} = 5\sigma_v^2 R / GM$ \citep{1992ApJ...Bertoldi}, provides a measure of the balance between a cloud’s internal kinetic energy and its gravitational binding energy, where $\sigma_v$ is the velocity dispersion, $R$ and $M$ are the cloud's effective radius and mass, respectively. Values below unity ($\alpha_\text{vir}< 1$) indicate gravitationally bound clouds, although a low virial parameter does not necessarily imply active collapse, as the collapse itself can increase the virial parameter again \citep[e.g,][]{2013ApJ...779..185K, 2021MNRAS.500.3027D}. The virial parameter is also sensitive to turbulence and stellar feedback processes such as ionisation and stellar winds, which can elevate $\sigma_v$ and hence increase $\alpha_\text{vir}$. The distributions of virial parameters, however, show much less separation between the two population of clouds: the associated clouds exhibit a slightly higher mean virial parameter (3.12) compared to the unassociated clouds (1.63), but the difference is much less pronounced than for the mass or surface density distributions. This result is not surprising because once massive star formation begins, feedback processes (e.g. turbulence, ionization) can raise the virial parameter, even if the initial collapse required $\alpha_\text{vir}< 1$, and we expect the feedback from \ion{H}{ii} regions or SNRs to disperse the natal molecular cloud, resulting in large virial parameters. To further test this, we divided the associated sample into gravitationally bound ($\alpha_\text{vir} < 1$) and unbound ($\alpha_\text{vir} > 1$) subsets. Within the sample of 90 associated clouds, 37/90 (41 per cent) clouds have $\alpha_\text{vir} < 1$ , and 53/90 (59 per cent) have $\alpha_\text{vir} > 1$, indicating no strong preference for bound versus unbound states among the associated clouds.

The FWHM of the $\alpha_\mathrm{vir}$ distributions are comparable, with the associated clouds showing only a marginally broader distribution suggesting less distinction in this physical property between the two cloud populations. A K-S test supports this observation, returning a p-value of 0.275, indicating that the virial parameter distributions may be drawn from the same parent distribution. Given the role of the virial parameter in star formation \cite{2013ApJ...779..185K}, this result suggests that, unlike mass, surface density, and linewidth, the virial parameter may not be as critical in differentiating the associated and unassociated clouds. 

We also inspected associated clouds with extreme $\alpha_\text{vir}\geq 10$. We identify 17 such associations, including 11 \ion{H}{ii} regions, 1 SNR, and 5 unclassified sources, with velocity dispersions in the range of 1.00--3.69 \kms\ and average gas surface densities in the range of 58--197 M$_\odot$\pc. These values suggest moderate velocity dispersions and a range of surface densities; the clouds' high $\alpha_\text{vir}$ likely reflects elevated velocity dispersion relative to mass or radius rather than uniformly low average gas surface density ($\Sigma$). Because most are associated with \ion{H}{ii} regions, this could indicate a larger internal motion of clouds, maybe partly caused by their active gravitational contraction, or by star-birth feedback, or sometimes both \citep{2021MNRAS.500.3027D}. They are promising case studies for targeted follow-up (e.g. expansion diagnostics, shock tracers) to test whether they are being disrupted or simply stirred.

\subsubsection{Scaling relations: Mass-size, Linewidth-size, $\sigma_\mathrm{v}^{2}$/R -Average gas surface density ($\Sigma$)}
\label{subsec:Scaling_relations}

\begin{table*}
\centering
\caption{Physical properties of SEDIGISM molecular clouds. 
         The columns, starting from left to right, stand for the following: the name of the interacting SMGPS extended source with SEDIGISM cloud, SMGPS source physical radius, radio luminosity ($L_r$) of interaction source, name of associated SEDIGISM molecular cloud, 
         High mass star formation tracer (HMSF), cloud adopted distance, cloud mass, cloud's deconvolved equivalent radius, cloud's average column density ($N_{\mathrm{H}_2}$), cloud's velocity dispersion ($\sigma_v$),  cloud's average gas surface density ($\Sigma$), cloud's virial parameter ($\alpha_\mathrm{vir}$), cloud star formation efficiency (SFE) and cloud dense gas fraction (DGF)}
\resizebox{\hsize}{!}{
\begin{tabular}{cccccccccccccccc}
\hline
\hline
SMGPS & R$_s$ & $L_{r}$ & SDG & HMSF & Dist. & Mass & 
$R_{dec}$ & $N_{\mathrm{H}_2}$ & Linewidth ($\sigma_{v}$)& $\Sigma$ & $\alpha_\text{vir}$ & SFE & DGF\\
(Source Name) & (pc) & (\WHz) & (Cloud Name) & - & (kpc) & (M$_\odot$) 
 & (pc) & ($cm^{-2}$) & (\kms) & (M$_\odot$\pc) & - &L$_\odot$/M$_\odot$  & -\\
\hline
S30009 & 40.58 & 5.18e+16 & SDG342.458+0.2648 & 0.0 & 7.35 & 3290 & 3.76 & 3.26e+21 & 1.270 & 74 & 2.14 & 0.05 & 0.29 \\
S30009 & 40.80 & 5.23e+16 & SDG342.270+0.2943 & 1.0 & 7.39 & 97560 & 13.51 & 7.58e+21 & 2.326 & 170 & 0.87 & 0.70 & 0.10 \\
S30009 & 43.89 & 6.06e+16 & SDG342.322+0.4698 & 0.0 & 7.95 & 900 & 1.95 & 3.18e+21 & 1.100 & 75 & 3.05 & -- & -- \\
S30009 & 48.53 & 7.40e+16 & SDG342.248+0.2545 & 0.0 & 8.79 & 890 & 2.02 & 2.90e+21 & 0.555 & 69 & 0.82 & -- & -- \\
S30009 & 39.80 & 4.98e+16 & SDG342.258+0.4208 & 0.0 & 7.21 & 200 & 1.06 & 2.15e+21 & 0.423 & 56 & 1.12 & -- & -- \\
S30009 & 47.70 & 7.15e+16 & SDG342.420+0.4232 & 0.0 & 8.64 & 21610 & 6.61 & 6.99e+21 & 1.865 & 158 & 1.24 & 0.53 & 0.11 \\
\hdashline
S30021 & 4.48 & 6.52e+14 & SDG341.209-0.1090 & 0.0 & 3.27 & 1330 & 2.62 & 2.72e+21 & 0.705 & 61 & 1.14 & -- & -- \\
S30021 & 4.48 & 6.52e+14 & SDG341.016-0.1252 & 0.0 & 3.27 & 6450 & 3.58 & 7.12e+21 & 1.052 & 160 & 0.71 & 0.12 & 0.14 \\
\hdashline
S30025 & 13.98 & 5.05e+15 & SDG340.936-0.0468 & 0.0 & 3.49 & 3030 & 3.29 & 3.96e+21 & 0.952 & 89 & 1.14 & 0.05 & 0.09 \\
S30025 & 13.62 & 4.80e+15 & SDG341.259-0.2767 & 1.0 & 3.40 & 29340 & 5.75 & 1.26e+22 & 1.254 & 283 & 0.36 & 2.44 & 0.33 \\
S30025 & 13.10 & 4.44e+15 & SDG341.209-0.1090 & 0.0 & 3.27 & 1330 & 2.62 & 2.72e+21 & 0.705 & 61 & 1.14 & -- & -- \\
S30025 & 12.54 & 4.07e+15 & SDG341.027+0.0083 & 0.0 & 3.13 & 580 & 1.30 & 4.76e+21 & 0.652 & 109 & 1.11 & 0.17 & 0.17 \\
S30025 & 13.10 & 4.44e+15 & SDG341.016-0.1252 & 0.0 & 3.27 & 6450 & 3.58 & 7.12e+21 & 1.052 & 160 & 0.71 & 0.12 & 0.14 \\
\hdashline
S30026 & 102.70 & 7.96e+16 & SDG341.499+0.2572 & 0.0 & 14.11 & 14620 & 8.05 & 3.17e+21 & 1.878 & 72 & 2.26 & -- & -- \\
S30026 & 101.75 & 7.81e+16 & SDG341.424+0.3949 & 0.0 & 13.98 & 30140 & 11.28 & 3.34e+21 & 2.542 & 75 & 2.81 & -- & -- \\
S30026 & 14.85 & 1.66e+15 & SDG341.307+0.1956 & 0.0 & 2.04 & 8560 & 4.30 & 6.55e+21 & 1.067 & 147 & 0.67 & 0.15 & 0.09 \\
S30026 & 99.57 & 7.48e+16 & SDG341.526+0.2338 & 0.0 & 13.68 & 23010 & 7.59 & 5.61e+21 & 1.374 & 127 & 0.72 & -- & -- \\
S30026 & 104.96 & 8.31e+16 & SDG341.674+0.2473 & 0.0 & 14.42 & 6420 & 4.84 & 3.77e+21 & 1.149 & 87 & 1.16 & 0.11 & 0.59 \\
S30026 & 12.52 & 1.18e+15 & SDG341.535+0.2136 & 0.0 & 1.72 & 20 & 0.38 & 1.87e+21 & 0.347 & 45 & 2.60 & -- & -- \\
S30026 & 12.08 & 1.10e+15 & SDG341.721+0.1069 & 0.0 & 1.66 & 1450 & 2.19 & 4.30e+21 & 1.003 & 97 & 1.76 & 0.89 & 0.04 \\
S30026 & 16.60 & 2.08e+15 & SDG341.686+0.1276 & 0.0 & 2.28 & 1300 & 2.16 & 3.94e+21 & 1.327 & 89 & 3.39 & -- & -- \\
S30026 & 104.52 & 8.25e+16 & SDG341.670+0.1709 & 0.0 & 14.36 & 10530 & 4.24 & 8.01e+21 & 1.731 & 187 & 1.40 & 1.17 & 0.27 \\
S30026 & 101.61 & 7.79e+16 & SDG341.511+0.0023 & 0.0 & 13.96 & 3260 & 2.88 & 5.16e+21 & 0.751 & 125 & 0.58 & -- & -- \\
\hdashline
S30037 & 18.10 & 4.80e+15 & SDG342.270+0.2943 & 1.0 & 7.39 & 97560 & 13.51 & 7.58e+21 & 2.326 & 170 & 0.87 & 0.70 & 0.10 \\
S30037 & 18.00 & 4.75e+15 & SDG342.458+0.2648 & 0.0 & 7.35 & 3290 & 3.76 & 3.26e+21 & 1.270 & 74 & 2.14 & 0.05 & 0.29 \\
S30037 & 19.47 & 5.56e+15 & SDG342.348+0.0705 & 0.0 & 7.95 & 7550 & 5.07 & 4.14e+21 & 1.789 & 94 & 2.50 & -- & -- \\
S30037 & 19.47 & 5.56e+15 & SDG342.457+0.1071 & 0.0 & 7.95 & 51350 & 14.57 & 3.43e+21 & 1.618 & 77 & 0.86 & 0.01 & 0.02 \\
\hdashline
S30058 & 5.94 & 1.19e+15 & SDG341.123-0.3523 & 1.0 & 3.23 & 7780 & 3.10 & 1.14e+22 & 1.048 & 257 & 0.51 & 0.87 & 0.10 \\
S30058 & 6.20 & 1.30e+15 & SDG341.235-0.3637 & 0.0 & 3.37 & 320 & 1.11 & 3.49e+21 & 1.642 & 81 & 11.08 & -- & -- \\
S30058 & 6.26 & 1.32e+15 & SDG341.259-0.2767 & 1.0 & 3.40 & 29340 & 5.75 & 1.26e+22 & 1.254 & 283 & 0.36 & 2.44 & 0.33 \\
\hdashline

$\vdots$ & $\vdots$ & $\vdots$ & $\vdots$ & $\vdots$ & $\vdots$ & $\vdots$ & $\vdots$ & $\vdots$ & $\vdots$ & $\vdots$ & $\vdots$ & $\vdots$ & $\vdots$ \\
\hline\hline
\end{tabular}
}
\begin{minipage}{18cm}
\vspace{0.2cm}
\footnotesize \textbf{Notes:} The source name is taken from the SMGPS extended sources catalogue by \citep{2025A&A...Bordiu}. The other quantities and physical properties are obtained
from the SEDIGISM catalogue by \citep{2021MNRAS.500.3050U} and \citep{2021MNRAS.500.3064S}, except for the columns of SMGPS source’s physical radius (which was
calculated from the angular radius of interacting source and distance from the associated cloud) and radio luminosity ($L_r$) which was determined by the source’s
flux density and distance from the associated cloud. Multiple clouds that are associated with same SMGPS source appear to be within similar cloud adopted distance range.
\end{minipage}

\label{tab:clouds' physical properties}
\end{table*}

\begin{figure*}
    \centering
    \includegraphics[width=\textwidth]{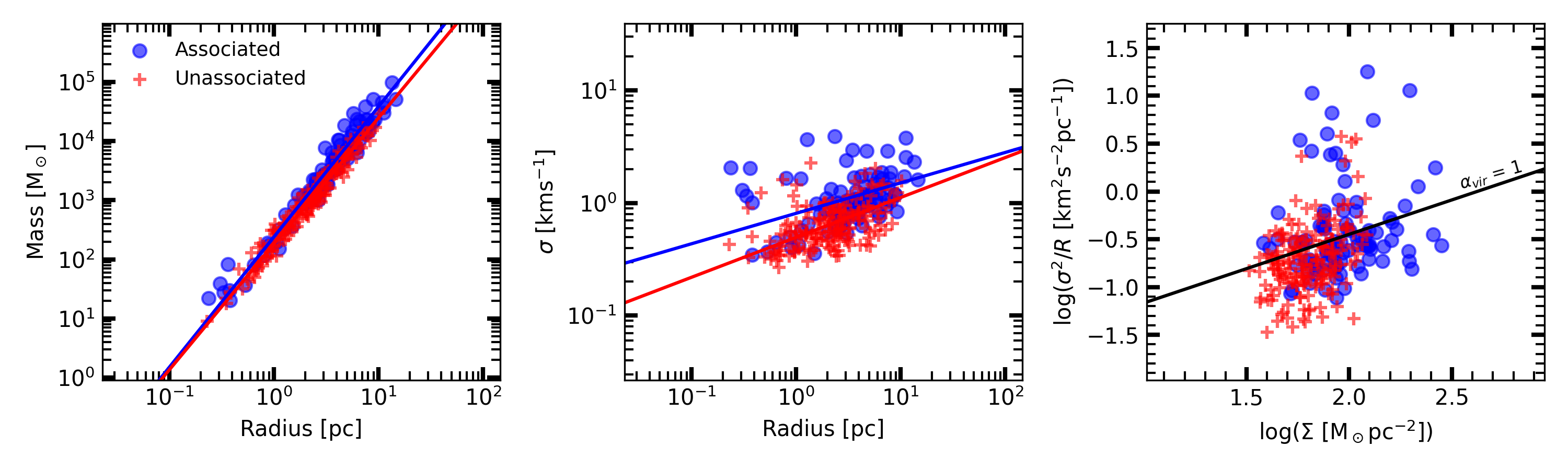}
    \caption{Larson's scaling relations of mass-size (left), Larson's first law \protect\citep{1981MNRAS.194..809L}: size-linewidth (middle), and \protect\cite{2009ApJ...699.1092H} relation $\sigma_\mathrm{v}^{2}/R$ versus $\Sigma$ (right) for the two categories of clouds; the associated SEDIGISM and unassociated molecular clouds. The blue and red continuous lines in the left and middle panels are linear fits of the associated and unassociated clouds, respectively. }
    \label{fig:Larson_law}
\end{figure*}

We present results of scaling relations of different physical properties of molecular clouds. Figure~\ref{fig:Larson_law} illustrates three scaling relations for both associated and unassociated SEDIGISM molecular clouds: mass versus size, linewidth (velocity dispersion) versus size, and the \cite{2009ApJ...699.1092H} relation between average gas surface density ($\Sigma$) and the squared linewidth to size ratio ($\sigma^{2}/R \propto \Sigma$). To examine the scaling relations between these physical properties of the clouds, we performed power-law fits using orthogonal distance regression (ODR). This method accounts for measurement uncertainties in both variables, making it more appropriate than ordinary least squares regression, which assumes all errors lie in the dependent variable. The ODR approach and the use of methods that consider errors in both axes have become common in studies of molecular cloud scaling relations  \citep[e.g.][]{2010A&A...519L...7L, 2012MNRAS.....Shetty,2013ApJ...779..185K, 2018MNRAS....Traficante}, especially where both mass and radius are observationally uncertain quantities.

For the mass-size relation, corresponding to the third Larson relation, the following results were obtained from ODR fits:
$M =10^{2.38\pm0.03}R^{2.20\pm0.05}$ for the associated clouds, and  $M =10^{2.26\pm0.01}R^{2.13\pm0.03}$ for the unassociated group.
The mass-size relationship shows that the masses of both associated and unassociated clouds scale nearly proportionally to their radii squared. The scatter around the fitted ODR relations is relatively small, suggesting that the relation holds consistently across both categories of clouds. The power-law exponents for this relation confirm that both groups generally follow a similar mass-size trend (also see Table \ref{tab:clouds' physical properties}) and that mass and size, given the uncertainties, are consistent with the power relation $M\propto R^{2}$ \citep{1981MNRAS.194..809L, 2010A&A...519L...7L, 2010ApJ...Kauffmann(a), 2010ApJ...Kauffmann(b), 2016ApJ...Nguyen-Luong}. However, associated clouds extend to higher masses (as observed from the intercepts--$10^{2.38}\approx 240\ \mathrm{M_\odot pc^{-2.20}}$ versus $10^{2.26}\approx 182\ \mathrm{M_\odot pc^{-2.13}}$), consistent with enhanced density and ongoing star formation activity.

In the middle panel of Figure \ref{fig:Larson_law}, the size-linewidth relation (Larson's first law) for the associated and unassociated populations are described by the ODR fitting results:
$\sigma_\mathrm{v} = 10^{-0.09\pm0.04}R^{0.27\pm0.06}$, and $\sigma_\mathrm{v} = 10^{-0.31\pm0.02}R^{0.35\pm0.04}$, respectively. Both relations exhibit slopes shallower than the canonical value of $\sim0.5$ derived by \cite{1987ApJ...319..730S}, and are closer to the original relation of \cite{1981MNRAS.194..809L}, who found a slope of $\sim0.38$. This implies a weaker dependence of linewidth on size compared to typical Galactic giant molecular clouds. Within the quoted uncertainties, the slopes of the two populations are consistent, with no statistically significant difference. The fitted intercepts however reveal that the velocity dispersions of the associated clouds are systematically larger for a given size compared to the unassociated clouds. This may reflect different dynamical conditions between the two groups. More than half ($\sim$59 per cent) of the associated clouds show $\sigma_v>1.0$ \kms, compared to only 15 per cent in the unassociated population. This high-dispersion tail likely reflects the impact of stellar feedback, as expected if the associated clouds are located near, or encompass, evolved \ion{H}{ii} regions and SNRs.
Nonetheless, the size-linewidth relation shows that both cloud types have a clear trend.

The right panel of Figure \ref{fig:Larson_law} depicts the so-called Heyer relationship between the logarithm of the average gas surface density ($\Sigma$) and the ratio of squared linewidth to size ($\log(\sigma^2/R$)). The Heyer relation tells us how the virial parameter distributions change as a function of surface density. The results, therefore, show that slightly more than half of the associated clouds (55 per cent) lie below the $\alpha_\mathrm{vir}$=1 line, while a greater fraction of the unassociated clouds fall below the line (57 per cent), and the associated clouds only tend to be more sub-virial at higher surface densities. These fractions falling below the line for both samples are not significantly different and it generally shows that there is no systematic trend of virial parameter with surface density. Interestingly, despite the spread on both axes of the associated sample being much larger, the $\alpha <1$ fractions are similar. Notably, the dispersion increases with mass and at least that the tail of high-virial parameter clouds is at the high surface density end of the distribution(s).

These distributions suggest that associated clouds tend to be more massive and exhibit higher velocity dispersions compared to unassociated clouds, reflecting their more active dynamical environments. The overall trend across the three scaling relations shows that the sample within this particular tile follows the scaling relations shown by \cite{2021MNRAS.500.3027D} for the full SEDIGISM sample, however, the associated and unassociated samples are barely distinguishable.

\subsubsection{SFE, DGF, surface density ($\Sigma$) versus radio luminosity, radius}
\label{subsec:Radio_luminosity}

\begin{figure}
    \centering
    \includegraphics[width=\columnwidth]{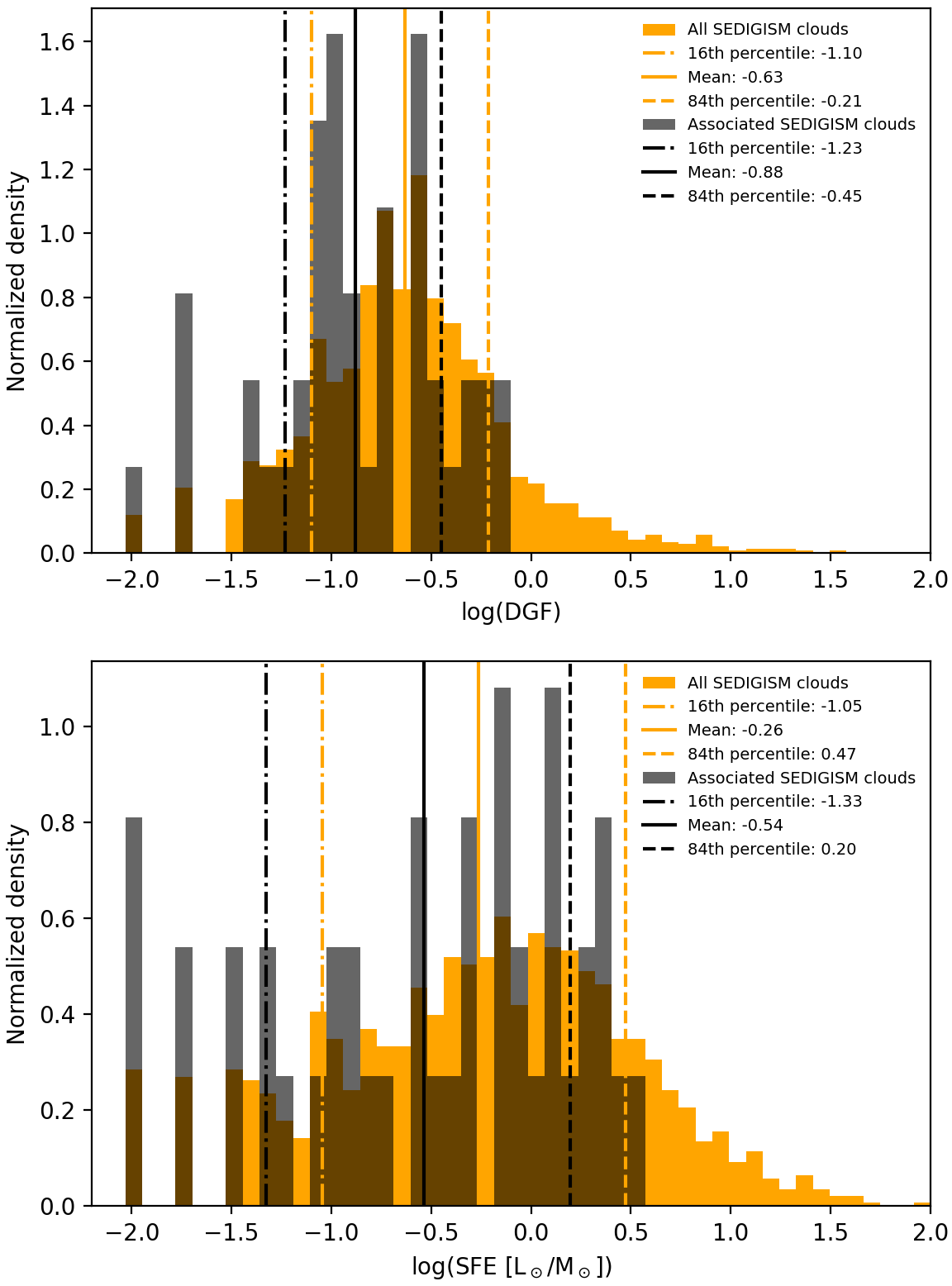}
    \caption{Top panel: Histogram of log of DGF of the whole SEDIGISM molecular clouds in the field (whole survey, orange) and associated SEDIGISM clouds (black). Bottom panel: Histogram of log of SFE (L$_\odot$/M$_\odot$) of the whole SEDIGISM clouds in the field (whole survey, orange) and associated clouds (black). Also shown are the legends of the mean, 16th and 84th percentiles for the DGF and SFE (L$_\odot$/M$_\odot$) for the whole SEDIGISM survey and of the associated SEDIGISM molecular clouds within G342 tile. Note that a small fraction (152/10663) of the SEDIGISM clouds exhibit DGF > 1 (log DGF > 0). As discussed by \citet{2018MNRAS.473.1059U, 2021MNRAS.500.3050U}, this arises from the large intrinsic uncertainties in the cloud and clump mass estimates, both of which were normalised to a common distance.}
    \label{fig:SFE_DGF_mean}
\end{figure}
 
In Figure ~\ref{fig:SFE_DGF_mean}, we show the distributions of SFE and DGF of the SEDIGISM molecular clouds. 22 per cent of all the SEDIGISM clouds in the field have an ATLASGAL counterpart, and therefore have a valid SFE and DGF measurement in \citet{2021MNRAS.500.3050U}, and 5 per cent have an associated HMSF tracer. In this work, we found that out of the 90 associated clouds, only 44 (49 per cent) clouds have both SFE and DGF measurements, of which only 12/44 (27 per cent) or 12/90 (13 per cent) have an associated HMSF tracer. These are higher fractions as compared to the overall rates for SEDIGISM clouds with the same properties in the field. From the 178 unassociated clouds, only 14/178 ($\sim$8 per cent) clouds have SFE and DGF measurements, and only 1/178 ($\sim$0.6 per cent) has an associated HMSF tracer. This shows that the clouds with associated SMGPS sources are more likely to be associated with ATLASGAL clumps. We therefore investigate whether we can identify any relationships between properties of the SMGPS sources themselves, such as radio luminosity or size, and the star-forming properties of the associated molecular clouds. For instance, high radio luminosity often indicates strong ionizing radiation or shock waves, which may compress nearby molecular clouds, potentially enhancing the cloud's star formation efficiency (SFE) and dense gas fraction (DGF) by driving gas into dense, star-forming cores. In addition, the physical size (i.e. radius) of an \ion{H}{ii} region or SNR might provide a useful proxy (under the assumption that such regions are generally expanding) for a timescale, and so we would like to test if any time-dependent relationships are apparent.

\begin{figure}
    \centering
    \includegraphics[width=\linewidth]{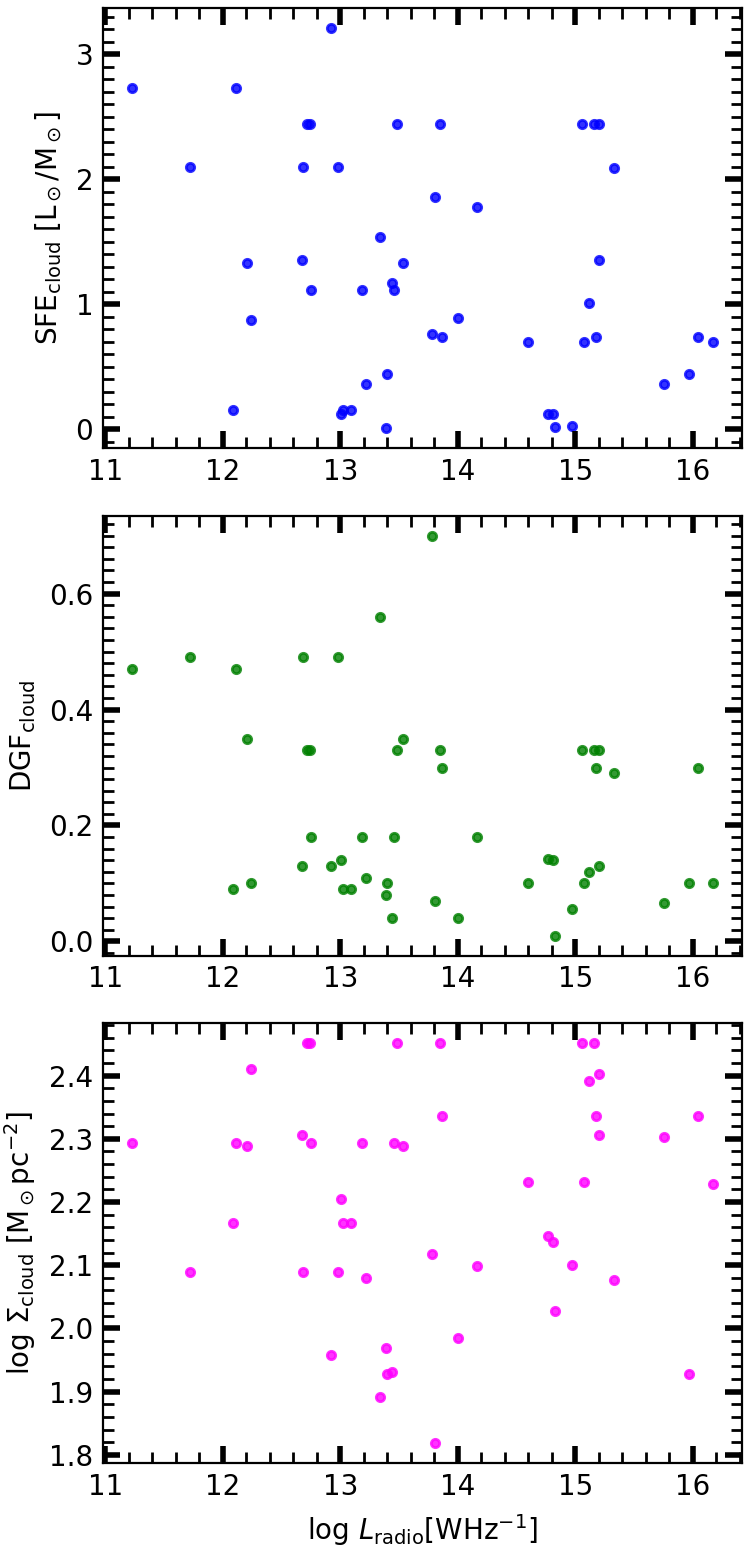}
    \caption{Relationship between radio luminosity of SMGPS extended \ion{H}{ii} regions (from the average best spectral window) and the associated best spectral window clouds' physical properties (mean values weighted by the SEDIGISM cloud brightness); Top: star formation efficiency, SFE in blue colour, Middle: dense gas fraction, DGF in green colour, and Bottom: average gas surface density, $\Sigma$ in magenta colour.}
    \label{fig:SFE_DGF_L}
\end{figure}

\begin{figure}
    \centering
    \includegraphics[width=\linewidth]{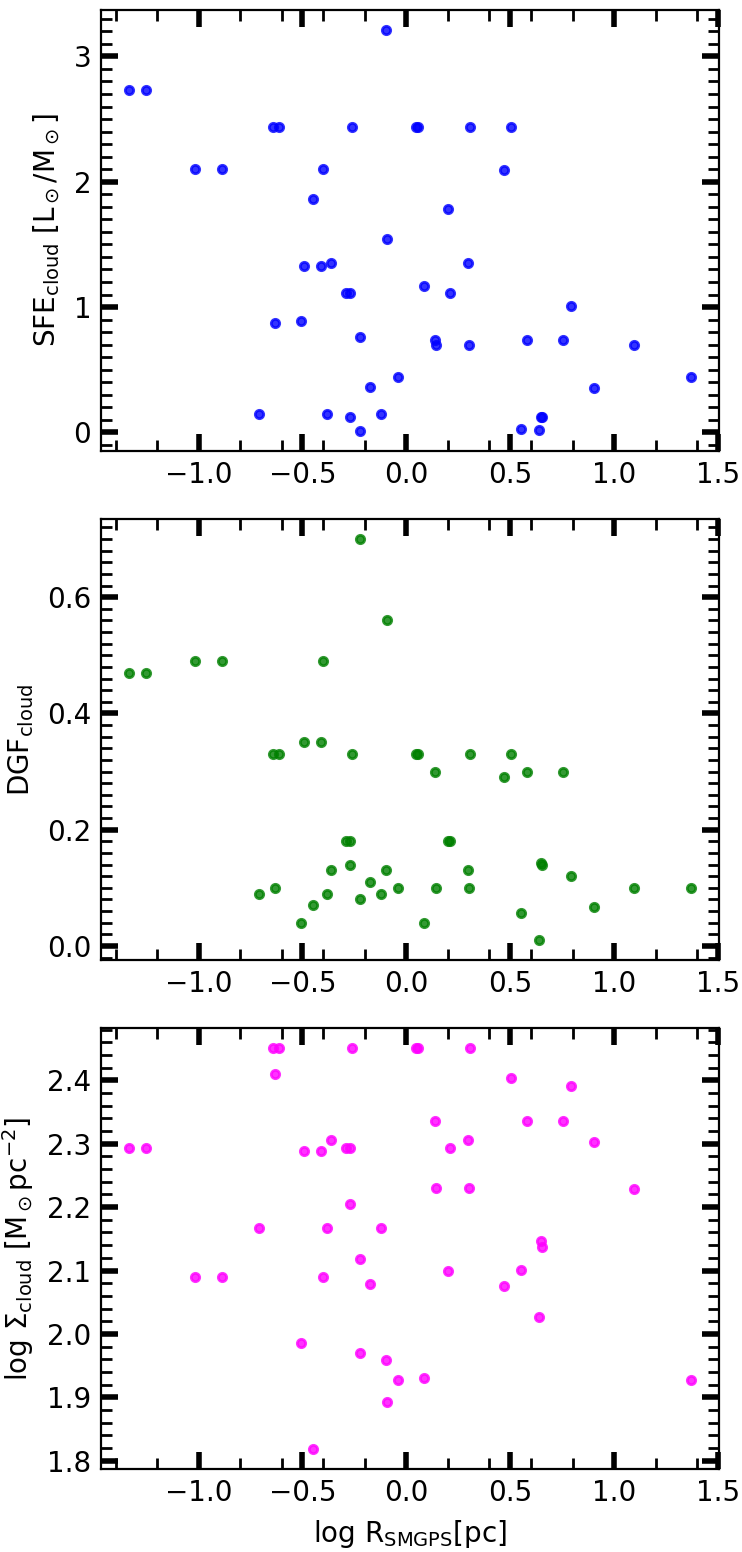}
    \caption{Relationship between size (radius in parsecs) of SMGPS extended \ion{H}{ii} regions (from the average best spectral window) and the associated best spectral window clouds' physical properties (mean values weighted by the SEDIGISM cloud brightness), and with the same figure arrangement and colours as in Figure \ref{fig:SFE_DGF_L}.}
    \label{fig:SFE_DGF_Radius}
\end{figure}

Figures \ref{fig:SFE_DGF_L} and \ref{fig:SFE_DGF_Radius} show the relationships between the luminosities and sizes of the SMGPS sources and the SFEs, DGFs, and surface densities of the clouds in the corresponding complexes. In cases where there are multiple clouds in a complex, the quantities presented are the mean values weighted by the clouds' brightness (i.e. total integrated intensity values). We see no relationships between the weighted mean SFE, DGF, or average gas surface density of the associated SEDIGISM molecular clouds with either the SMGPS luminosity or radius, suggesting that while feedback mechanisms might play a role in shaping SFE and DGF, other factors are likely more influential, including limitations in our matching method. Another reason for the results is possibly due to the limited SEDIGISM data coverage (only a single tile analysed so far as part of this pilot study). With a larger dataset covering more molecular clouds, stronger correlations might emerge, providing a clearer understanding of how feedback mechanisms from massive stars influence SFE and DGF on a Galactic scale.

\section{Discussion}
\label{sect: Discussion}
\subsection{The assignment of CO-based velocities to radio sources}
\label{subsec: confidence_method}
In Figure \ref{fig:SMGPS_WISE} we compared the velocities of WISE \ion{H}{ii} regions, derived from radio recombination lines, to the velocities of SMGPS \ion{H}{ii} regions that we derived from the associated SEDIGISM CO emission. The radio recombination line (RRL) velocities serve as an independent validation of our method of associating SMGPS sources with SEDIGISM clouds. There is a very strong linear relationship observed, and the average difference between the two categories of velocities is $\sim$ 6\,\kms. This statistical difference is reasonable when considering the fact that these velocities are derived differently since the WISE RRL velocities are derived directly from the ionized gas in \ion{H}{ii} regions, while the SEDIGISM CO velocities are associated with molecular gas in the surrounding environment. The physical separation between the ionized and molecular components, even within the same region, can naturally lead to slight velocity offsets. Overall, the linear relation and small average velocity difference implies that the SMGPS CO-derived velocities align closely with those from WISE, indicating a high degree of reliability in our velocity measurements. We therefore consider that for our threshold emission fraction (\fw$\geq 0.3$) SMGPS-SEDIGISM associations, the CO velocities are suitably robust, and can be applied for the full sample, including the vast majority which do not have RRL measurements.

\subsection{Supernova remnants}
\label{subsec: SNRs}
\begin{figure*}
    \centering
    \includegraphics[width=\textwidth]{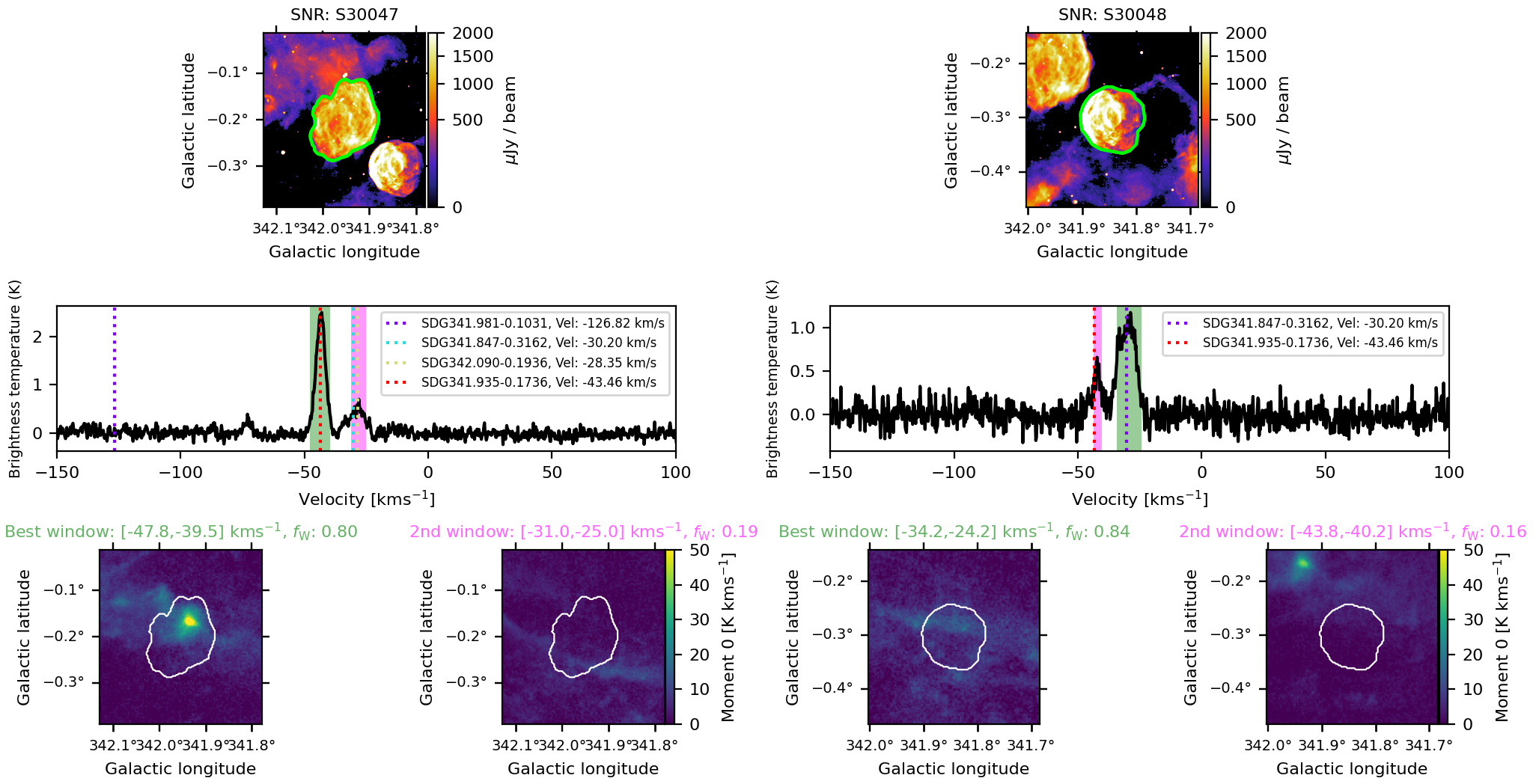}
    \caption{Multiwavelength, spectral and CO analysis of the SNRs S30047 (left) and S30048 (right). Top row: SMGPS 1.3 GHz continuum images, with the green contours outlining the SMGPS source (SNR) boundaries. Middle row: Average $^{13}$CO(2--1) spectra extracted from the SEDIGISM survey toward each SNR. Vertical dashed lines mark the centroid velocities of SEDIGISM molecular clouds intersecting the SNRs, while shaded regions indicate the velocity windows considered. Bottom row: Integrated $^{13}$CO moment-0 maps corresponding to the selected velocity ranges. For each SNR, the best-matching velocity window (left panel, green label) and the second best velocity window (right panel, magenta label) are shown, with the white contours tracing the SNR boundary. The emission fraction (\fw) for each velocity window is given in parentheses.}
    \label{fig:SNRs_plots}
\end{figure*}

Among the SMGPS extended sources associated with SEDIGISM molecular clouds are two SNRs, S30047 (G341.953-00.202) and S30048 (G341.844-00.304), which have radii of 6.87 arcmin and 5.15 arcmin, respectively, and which are found in close proximity being separated by only $\sim0.1\degr$. In both cases, the $^{13}$CO (2--1) zeroth moment maps (Figure \ref{fig:SNRs_plots}, bottom row) corresponding to the best velocity windows reveal molecular structures that overlap spatially with the SNR boundaries traced by the SMGPS continuum. For S30047, the brightest $^{13}$CO emission is concentrated towards the higher latitude edge of the remnant, a high column density region located within the SNR boundary and shows a reasonable morphological correspondence, though without clear evidence of compression; the match is stronger than for the alternative velocity window. For S30048, CO emission is more diffuse, indicative of a weaker or much less convincing morphological match but still stronger than the alternative velocity window. These associations highlight that the morphology of the continuum emission provides a useful guide to identifying the molecular components most likely affected by the expanding SNRs.
 
In Figure \ref{fig:SNRs_plots} (middle row), we show the mean $^{13}$CO (2--1) spectra, averaged over the SMGPS area, which two main velocity components separated by $\sim$13\,\kms, suggesting that the two remnants may be interacting with (or have originated from) different parts of a common molecular complex. However, we maintain that only the clouds falling within the best velocity windows (shaded green) are considered physically associated with the SNRs, and we adopt their systemic velocities and kinematic distances as those of the remnants. For the SNR S30047, the best-matching velocity window (-47.8 to -39.5\kms) has an emission fraction (\fw) of 0.804 with the associated cloud (SDG341.935-0.1736) with a systemic velocity of -43.7\,\kms. The cloud has an adopted distance of 3.4\,kpc and is $\sim 6.12$ pc in physical size (radius), has HMSF tracers, with velocity dispersion $\sigma_v=1.71$\,\kms, and virial parameter $\alpha_\text{vir}=1.1$, consistent with a bound, and dense environment. We therefore estimate the distance of this SNR to be 3.4 $\pm$ 0.5 kpc based on the final adopted kinematic distance of the cloud \citep{2021MNRAS.500.3027D}. Similarly, SNR S30048 best matches cloud SDG341.847-0.3162 at a centroid velocity of -30.2\,\kms\ with a moderately higher emission fraction (\fw) of 0.84 for the best-matching velocity window (-34.2 to -24.2\,\kms). The cloud has relatively small radius of 3.48 pc, a high velocity dispersion of 2.96\kms, and a large virial parameter of 10.84 (also see Table \ref{tab:extreme alpha cases}), indicating that it is turbulent and gravitationally unbound. We thus estimate the kinematic distance of SNR S30048 to be 2.6 $\pm$ 0.6 kpc based on the associated cloud distance from \cite{2021MNRAS.500.3027D}.

These are the first reliable kinematic distance estimates for the two SNRs based on their associations with molecular clouds. \cite{2022ApJ...Ranasinghe} presented kinematic distance estimates of 15.8 $\pm$ 0.6 kpc for both SNRs, using hydroxyl radical (OH) emission described in \citet{1998AJ....Koralesky}. However, neither of these SNRs were detected interferometrically by \citet{1998AJ....Koralesky} and the $v_{\rm lsr}$ values presented refer to the central velocities of the observational setup, \emph{not} the velocity of any OH emission. \citet{1997AJ....Green} refer to a single dish detection of OH for these SNRs but do not give the  $v_{\rm lsr}$ for the detection. In contrast, our $^{13}$CO (2--1)-based method establishes the first systemic velocities for these SNRs and hence a more reliable kinematic distance estimate. Importantly, adopting the larger maser-derived distance of 15.8 kpc would imply physical radii of $\sim$30 pc for S30047 and $\sim$23 pc for S30048, compared to the much more reasonable $\sim 6.9$ pc and $\sim 3.8$ pc obtained at our adopted distances. Such large radii would correspond to old, evolved remnants inconsistent with their observed morphologies.

If the progenitors of S30047 and S30048 were core–collapse supernovae, they must have formed in molecular clouds within the past few 10 Myr. This is because, according to \cite{2018ApJS...Limongi}, the main–sequence lifetime of the least massive supernova progenitors ($\sim8\mathrm{M_\odot}$) is $\leq 30-40$ Myr, as most massive stars take upwards of 10 Myr to explode. Even those with mass ($>40\mathrm{M_\odot}$) do not explode before $\sim4$ Myr \citep{2006ApJ...Limongi}, and cluster simulations confirm that no supernovae occur before $\sim5$ Myr \citep{2023MNRAS....Parker, 2024ApJ...Eatson}. These timescales imply that massive stars generally remain near their birth environments until explosion, but dynamical ejection can allow some to travel tens to hundreds of parsecs before the supernova, weakening or erasing any molecular counterpart. Within this framework, the relatively convincing match for S30047 is consistent with a progenitor that exploded close to its parent environment, whereas the weaker correspondence for S30048 may still be compatible with a core–collapse origin if the progenitor had drifted away from its birth site. Thus, our results not only resolve the velocity controversy for these SNRs but also highlight how molecular cloud associations can provide both kinematic distances and environmental context for understanding SNR evolution. Our pilot study thus highlights the potential of associating larger datasets of SMGPS extended sources and SEDIGISM clouds, building more comprehensive catalogues of SNR distances and their surrounding environments.

\subsection{The co-evolution of radio, CO emission, and \ion{H}{ii} regions}

Our results suggest a strong coupling between the radio emission of the SMGPS, the molecular CO emission of SEDIGISM, and the evolutionary state of \ion{H}{ii} regions. 
In the top panel of Figure~\ref{fig:L_radio vs mass}, we find that radio emission correlates with the molecular mass of the complexes, indicating that the strength of the ionized emission is coupled to the available gas reservoir. This is consistent with studies showing that compact and ultra-compact \ion{H}{ii} regions are preferentially embedded in massive molecular clumps, and that clump mass correlates with the properties of embedded star or cluster properties \citep[e.g.][]{2013MNRAS.435..Urquhart}. Comparable relations appear in the literature across scales, with radio luminosity found to be correlated to envelope mass of protostellar cores \citep{2011MNRAS...AMI}, and the correlation of radio luminosity to CO emission on the scale of galaxies \citep{2002A&A...Murgia}. Our results therefore form part of a broader, multi-scale trend linking molecular gas to radio luminosity across scales.

The middle and bottom panels of Figure~\ref{fig:L_radio vs mass} show correlations with respect to SMGPS source size, which is a reasonable proxy for evolutionary state. We find a positive correlation between SMGPS source angular size and \lm\ ratio, which we interpret as a signature of more efficient conversion of molecular to ionized emission with the evolution and expansion of \ion{H}{ii} regions. 
Comparable trends are reported by the ALMA Three-millimeter Observations of Massive Star-forming regions (ATOMS) survey \citep{2024MNRAS....Zhang}, which finds that the most massive and luminous clumps host the youngest and most compact H{\sc ii} regions, while more evolved and extended regions are embedded in more dispersed gas environments. At the same time, the fraction of CO emission overlapping with SMGPS sources is seen in the bottom panel of  Figure~\ref{fig:L_radio vs mass} to decrease with increasing SMGPS source size, consistent with results from the Physics at High Angular resolution in Nearby GalaxieS (PHANGS) survey on galaxy scales that show diminished molecular overlap in evolved regions \citep{2023A&A...Zakardjian}. These trends point to a scenario in which stellar feedback progressively disrupts and disperses natal molecular clouds \citep[e.g.][]{2022...Chevance}.
Furthermore, pixel-by-pixel comparison of the SMGPS radio intensity to the SEDIGISM $^{13}$CO$(2-1)$ zeroth moment intensity show a strong positive correlations of the same strength between these quantities for both an extended \ion{H}{ii} region (see Figure~\ref{fig:complex_pixel-pixel_scatter}) and a compact \ion{H}{ii} region (see Figure~\ref{fig:compact_pixel-pixel_scatter}), again suggesting the co-evolution of ionized and molecular emission across evolutionary stage.

The case studies of SMGPS sources, S30009 and S30035, provide further insight into how feedback and molecular structure may interact. In the case of S30009 (Figure \ref{fig:complex_pixel-pixel_scatter}), the large and clumpy CO distribution, broadened velocity dispersions and well-developed ionized cavity point to an evolved region where massive stellar feedback has already reshaped the molecular morphology. The observed correlation slope implies that as the SMGPS intensity increases (or the ionized gas expands), the SEDIGISM CO emission continues to proportionally participate in the process, either through compression at the ionization front or disruption through turbulence.  
While such behaviour is often interpreted in the framework of “triggered” star formation \citep[e.g.][]{1998ASPC..148..150E, 2005A&A...433..565D, 2006A&A...446..171Z, 2008ASPC..Deharveng}, we emphasise that our data do not allow us to establish a causal link. In particular, it remains unclear whether new star formation here would not have occurred in the absence of the expanding \ion{H}{ii} region, or whether the \ion{H}{ii} region is primarily excavating and unveiling a pre-existing dense ridge. The morphology of the SMGPS continuum emission shows a network of dense ridges, or pillars, that radially point towards the central cavity, with the CO emission aligned along them. Such a configuration is difficult to explain solely as a product of feedback-driven compression. Instead, it may be indicative of a pre-existing hub-filament system (HFS). HFSs are networks of spatially converging filaments \citep{2009ApJ...Myers}, with ordered velocity gradients along their lengths indicating gas inflow towards the hub centre where active star formation is promoted \citep[e.g.][]{2013A&A...Peretto, 2014A&A...Peretto, 2018A&A...Gwen, 2019A&A...Trevino, 2023MNRAS..Liu}. In the picture of HFS evolution presented by \cite{2020A&A...Kumar}, an \ion{H}{ii} region at the hub centre would be expected to erode the dense hub at it expands, leaving behind dense hub-filament ridges as pillars that point radially towards the cavity centre. One would also expect the previously ordered velocity gradient to become disordered, coupled with a broadening of the velocity dispersion. A similar transition from dense hub–filament system to an \ion{H}{ii} region morphology was previously documented in SDC13 prototypical hub by \cite{2014A&A...Peretto} and \cite{2018A&A...Gwen}, which also showed evidence of fragmentation along the filaments. This evolved picture is consistent with both our morphological and kinematic results towards S30009, therefore we suggest that S30009 could be an example of an evolved HFS undergoing erosion. By contrast, the compact S30035 shows a more centrally concentrated morphology and relatively uniform kinematics, consistent with an earlier evolutionary stage in which the molecular gas remains intact and the ionization is still localised.

Taken together, our results indicate that the radio continuum, CO emission, and \ion{H}{ii} regions co-evolve as massive stars interact with their parent molecular clouds. The consistent scaling between the radio and CO emission, across both compact and extended \ion{H}{ii} region morphologies, supports this view across evolutionary stage. 

\subsection{Statistical and Scaling tests}
\label{subsec: statistical_tests}
The histograms in Figure \ref{fig:mass} show that the masses, average surface densities, and linewidths (first, second and third panels from left, respectively) of the associated SEDIGISM molecular clouds extend to values approximately an order of magnitude greater than those of the unassociated sample. In contrast, the distributions of virial parameter (right panel of Figure \ref{fig:mass}) show no clear difference between the two populations. Nevertheless, some associated clouds show particularly large virial parameters ($\alpha_\mathrm{vir}\geq10$, see Figure \ref{fig:Larson_law} to the right-most, and also Table \ref{tab:extreme alpha cases} ). These outliers are interesting candidates for clouds potentially disrupted or strongly affected by feedback, and merit closer examination in follow-up work. This particular case of physically affected clouds with high virial parameters is explained by \cite{2022A&A...664A..84N} that it could be possible that the majority are ring-type clouds showing signs of expansion and could be potential sites for future massive star formation. Nevertheless, in our case, the general comparison between the associated and unassociated SEDIGISM molecular clouds in terms of the distribution of the virial parameters (see Figure \ref{fig:mass} to the right-most) shows no clear contrast and the p-value from the K-S test in Section \ref{subsec: hist_distribution} suggests that there is no statistical significance. This lack of contrast is perhaps unsurprising, since once massive star formation begins, the kinetic energy from the collapse itself, as well as feedback can increase linewidths and hence virial parameter, even in regions that were originally gravitationally stable. 

In this work, we have shown that almost half (49 per cent) of the associated clouds have measurable SFE and DGF values i.e. are associated with dense clumps in ATLASGAL. More than a quarter (27 per cent) of these associated clouds with SFE and DGF values are likely connected to high-mass star formation. The mean and 84th percentile SFE values -0.54 (L$_\odot$/M$_\odot$) and 0.20 (L$_\odot$/M$_\odot$), and DGF values -0.88 and -0.45, respectively, shown in the associated clouds in Figure \ref{fig:SFE_DGF_mean}, suggest that a significant proportion of these clouds contain relatively dense, bound gas regions, potentially undergoing collapse or active star formation. This supports the idea that the association with active \ion{H}{ii} regions or supernova remnants, for instance, the SMGPS extended sources, may promote denser gas concentration, hence increasing the likelihood of triggered star formation \citep[e.g.,][]{2005MNRAS...Dale, 2007MNRAS...Dale, 2016ApJ...Kendrew, 2022A&A...663A..56N, 2022A&A...664A..84N}. On the other hand, only one unassociated cloud has associated HMSF tracers, and less than 8 per cent (14/178) of them display measurable SFE and DGF values. Note, however, that while this means that only a very small fraction of unassociated clouds have ATLASGAL clumps within them, this does not mean they are devoid of all star formation activity--they might host low mass star-formation, or simply be in earlier stages of evolution. Therefore, it is a good indication that having SFE and DGF values alone may not fully describe a cloud's star-formation potential but other physical properties may also play key roles and must be explored as well. This is consistent with the need for a broader approach to assessing star-forming potential in molecular clouds, as highlighted in recent studies \citep[e.g.,][]{2022A&A...663A..56N, 2022A&A...664A..84N}. Our use of HMSF tracers should be interpreted with care. We emphasise that we followed the approach of \cite{2021MNRAS.500.3027D}, where the HMSF property is purely derived from ATLASGAL cross-matches. Therefore, we do not claim evidence for triggered star formation in individual clouds. Instead, our aim is to use the combination of HMSF tracers, SFE, and DGF as statistical diagnostics of the star formation properties of associated versus unassociated SEDIGISM molecular clouds. This approach is consistent with the cautionary notes of \cite{2021MNRAS.500.3027D} and \cite{2017A&A...601A.124S} and follows the framework of \cite{2021MNRAS.500.3050U} to provide a population-level view of massive star formation rather than a case-by-case analysis of triggering.

In Section \ref{subsec:Scaling_relations}, we examined the mass--radius, size--linewidth relationships (i.e. Larson's Laws) and the Heyer relation. The size--linewidth scaling relation is approximated as $\sigma_\mathrm{v} \sim R^{0.27\pm0.06}$ for the associated clouds and $\sigma_\mathrm{v} \sim R^{0.35\pm0.04}$ for the unassociated clouds (see Figure \ref{fig:Larson_law}, middle panel). \cite{1981MNRAS.194..809L} reported $\sigma_\mathrm{v} \sim R^{0.38}$ for size-velocity dispersion correlation of Galactic molecular clouds in general, implying a stronger correlation between linewidth and size. Within the uncertainties, the slopes of our associated and unassociated populations are statistically consistent, and therefore no significant difference in scaling can be established. More robust, however, is the systematic offset in linewidths: the associated clouds exhibit higher velocity dispersions, with a mean $\sigma_v\approx1.26$ \kms, compared to $\approx0.71$ \kms\ for the unassociated clouds. This indicates that although both populations follow similarly shallow linewidth–size relations relative to Larson’s law, the associated clouds maintain consistently larger velocity dispersions at a given radius, likely reflecting enhanced turbulence driven by stellar feedback, while the unassociated clouds may be influenced by additional factors that disrupt this scaling.

Generally, since we have a good mass-size relationship and a good scatter in the size-linewidth correlation among the associated molecular clouds, these findings advance previous studies \citep[e.g.,][]{2019ApJ...885...50W,2022A&A...663A..56N, 2022A&A...664A..84N} by emphasizing the significance of high-mass, high-density clouds as potential sites for future star formation. The interpretation would be that the increased linewidths probably reflect expanding motions in the clouds induced by the expanding \ion{H}{ii} regions and that more massive clouds with broader velocity dispersions are actively triggered star formation sites \citep{2019ApJ...885...50W}. Our analysis suggests that a majority of associated clouds are dynamically distinct from their unassociated counterparts, particularly in regions with turbulent, high-velocity dispersions.

\section{Conclusions}
\label{sec: Conclusion}

In this paper, we have presented a pilot study examining the spatial and kinematic relationships between molecular clouds, and \ion{H}{ii} regions and supernova remnants (SNRs) in a slice of the Milky Way centred on $\ell = 342.5^\circ$, $b=0^\circ$. Using data from the radio survey SMGPS and the SEDIGISM survey of $^{13}$CO (2--1) emission, we identified positional and velocity associations between radio continuum sources and molecular clouds to characterize their physical connections and the influence of feedback processes. We used $^{13}$CO (2--1) spectra averaged over the SMGPS source footprints to make associations with SEDIGISM clouds, and adopted their radial velocities. 
This approach allowed us to effectively mitigate uncertainties in velocity associations. Many sources are grouped in highly confused lines of sight, and the method yielded multiple plausible velocities to which we assigned a low emission fraction. 

This work, from the SEDIGISM tile (G342), has revealed that 159 of the 268 ($\sim$60 per cent) SEDIGISM molecular clouds in the field intersect 131 SMGPS sources along the line of sight. Of these, approximately 90 molecular clouds across the Galactic disc, (34 per cent) fall within the best-matching velocity windows and are therefore considered associated, forming 131 molecular–radio complexes comprising 57 \ion{H}{ii} regions, 2 SNRs, and 72 unclassified sources. This suggests that a reasonable fraction of the molecular clouds are interacting with \ion{H}{ii} regions or SNRs, presenting from a moderately to highly dynamic view of star formation. We also validated our association method for deriving velocities to the SMGPS sources by comparing the sample with the 20 available WISE \ion{H}{ii} regions that have direct measurements of velocities from radio recombination lines (RRLs). The CO-derived velocities of our high-emission fraction SMGPS sample are closely aligned with the RRL velocities, with an average discrepancy of $\sim$6\,\kms, and median absolute velocity difference of 2.96 \kms, giving confidence in our CO-derived velocities across the full sample. Our main findings are as follows:

\begin{enumerate}[leftmargin=*, nosep]

    \item The clouds associated with \ion{H}{ii} regions and SNRs exhibit distinct physical properties compared to unassociated clouds. The molecular clouds associated with SMGPS sources are, on average, more massive, and have higher surface densities than unassociated counterparts. The virial parameter distributions are similar, which may reflect a balance between gravitational collapse of massive star-forming clouds, and expansion of \ion{H}{ii} regions. However, we see the associated clouds have a small tail with high virial parameters, which may corresponding to clouds under the process of dispersal driven by the expansion of \ion{H}{ii} regions \citep{2021A&A...645A.110Y}.

    \item The correlation between \lradio\ and \mcomplex\ for the associated complexes provides evidence that radio luminosity and molecular mass scale together, and that evolutionary effects are imprinted on such relations. Specifically, we find that \lm\  increases with the angular size of SMGPS sources, while pixel coverage decreases with physical size. This indicates that as \ion{H}{ii} regions (and SNRs) expand, they become more luminous relative to their natal gas reservoir while progressively eroding their molecular environments. These trends reinforce the picture of co-evolution between ionized and molecular gas, with feedback leaving measurable imprints on the surrounding molecular clouds.
    
    \item We find that associated SEDIGISM molecular clouds follow Larson's law \citep{1981MNRAS.194..809L} for mass--size and not exactly with size--linewidth scaling relations. The associated sample seems to have an elevated velocity dispersion, a reflection of expanding motions in the clouds induced by the expanding \ion{H}{ii} regions. However, mass--size relations remain similar or consistent across the two categories of clouds.

    \item By associating SMGPS extended sources with SEDIGISM clouds, we derived new kinematic distances for the two SNRs; S30047 (G341.953-00.202) and S30048 (G341.844-00.304) in the field, and have revised the distances from the 15.8 $\pm$ 0.6 kpc \citep{2022ApJ...Ranasinghe} to 3.4 $\pm$ 0.5 and 2.6 $\pm$ 0.6 kpc, respectively. This highlights the potential for similar distance  measurements to SNRs in the rest of the SMGPS catalogue.

    \item The associated SEDIGISM clouds do not show significant correlations between the radio luminosity or physical size of SMGPS sources with associated clouds' SFE, DGF, or average gas surface density ($\Sigma$), whose relationship might be expected to show evidence of triggered star formation. However, it is possible that the sample in this pilot study is simply too small to identify such a signal.
\end{enumerate}

This work establishes a framework for a large-scale study of triggered star formation processes across the Galactic plane, highlighting the need to refine our understanding of cloud dynamics, associations, and feedback mechanisms. By extending this study across the full overlap between SMGPS and SEDIGISM in future, we may dramatically increase the sample size, and the strength with which such conclusions might be drawn.

\section*{Acknowledgements}
We thank the anonymous referee for thorough and constructive comments that substantially improved this paper. Both the SEDIGISM data acquired with the Atacama Pathfinder EXperiment (APEX) and the MeerKAT data obtained from the MeerKAT telescope are primarily used in this study. APEX is a collaboration among the Max-Planck-Institut fur Radioastronomie, the European
Southern Observatory, and the Onsala Space Observatory. MeerKAT telescope is operated by the South African Radio Astronomy Observatory, which is a facility of the National Research Foundation, an agency of the Department of Science and Innovation. This research would not have been possible without the Astropy project and the NASA ADS.
We acknowledge and thank the Development in Africa with Radio Astronomy (DARA) project for supporting this research through the UK’s Science and
Technologies Facilities Council (STFC) grant ST/Y006100/1.  MAT acknowledges support from STFC grant awards ST/R000905/1 and
ST/W00125X/1. MOL extends his gratitude to the University of Leeds for their support during his visits.

\section*{Data Availability}

This paper uses data from:\\
The SEDIGISM survey which includes projects 092.F-9315 and 193.C-0584, and the processed data products are available at \url{https://sedigism.mpifr-bonn.mpg.de/index.html}, which was constructed by James Urquhart and hosted by the Max Planck Institute for Radio Astronomy.\\
The SMGPS survey data \citep{2024MNRAS.531..649G} and the SMGPS Extended Source Catalogue \citep{2025A&A...Bordiu} are available at \url{https://doi.org/10.48479/3wfd-e270}, and \url{https://doi.org/10.48479/t1ya-na33} respectively. The Extended Source Catalogue is also available at the CDS via anonymous ftp to \url{https://cdsarc.cds.unistra.fr/viz-bin/cat/J/A+A/695/A144} or via \url{http://cdsweb.u-strasbg.fr/cgi-bin/qcat?J/A+A/}.\\
The data from the Wide-field Infrared Survey Explorer (WISE) Catalogue of Galactic \ion{H}{ii} Regions \citep[][]{2014ApJS..Anderson(2), 2018ApJS..Anderson(3)} are available at \url{http://astro.phys.wvu.edu/wise}.\\
We will provide an electronic version of Table \ref{tab: merged_catalogue} alongside this manuscript.



\bibliographystyle{mnras}
\bibliography{paper} 




\appendix

\section{Evaluation of the strength of SMGPS-SEDIGISM association by weighted standard deviations of cloud's systemic velocity}
\label{sec:complex_matches}
\begin{figure}
    \centering
    \includegraphics[width=\linewidth, height=\linewidth]{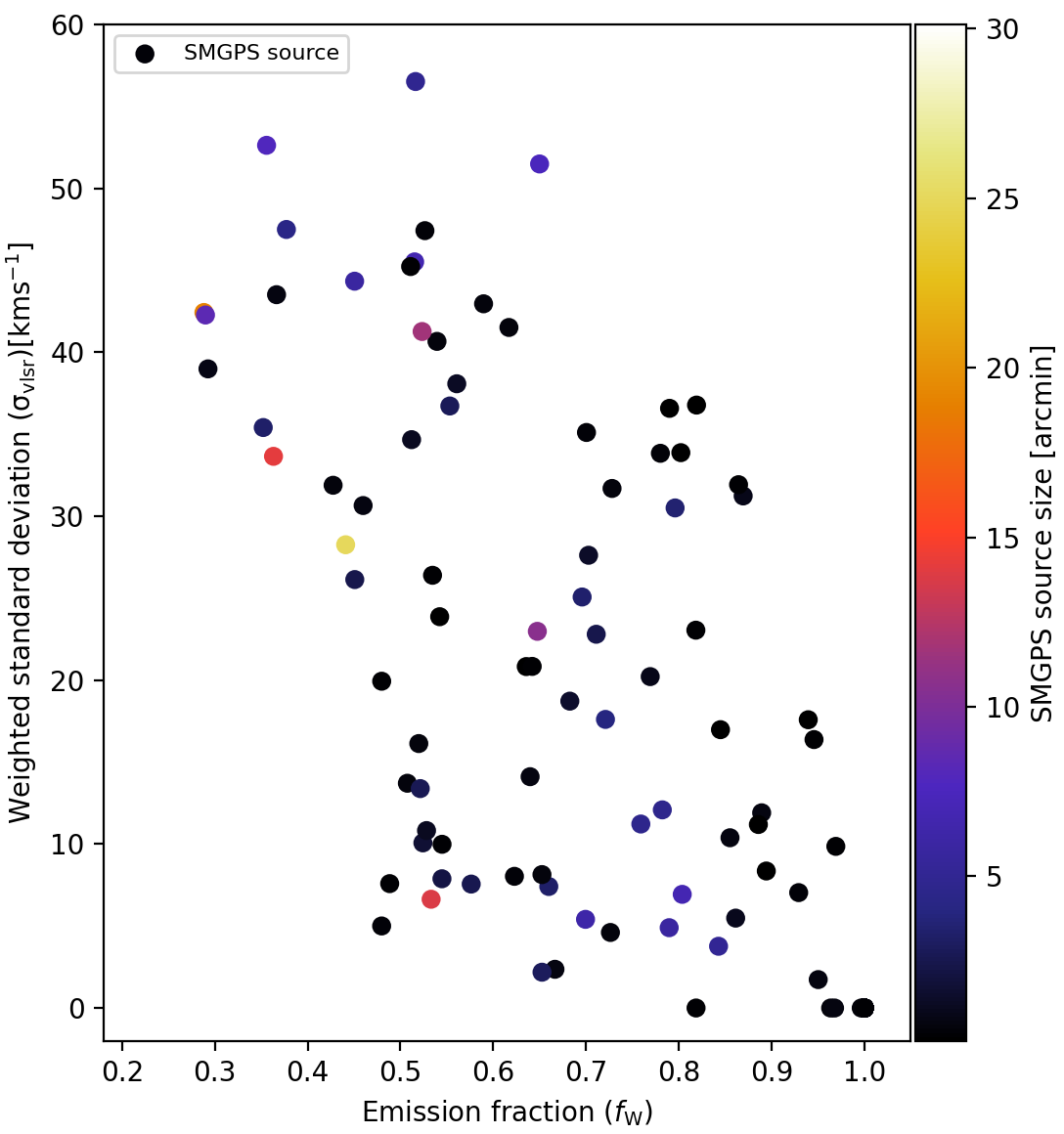}
    \caption{Window's emission fraction (\fw) as a function of $^{13}$CO brightness-weighted standard deviation ($\sigma_{v_\text{lsr}}$) plot of the interacting SMGPS sources and the associated SEDIGISM molecular clouds within the best-matching velocity windows. The colour bar represents the angular sizes in arcminutes of the SMGPS sources.}
    \label{fig:weighted_std_confidence}
\end{figure}

To evaluate the strength of the associations between SMGPS sources and SEDIGISM molecular clouds, we analyzed how the emission fraction (\fw) values of these clouds and sources correlate with the weighted standard deviation of their centroid velocities ($\sigma_{v_\text{lsr}}$), considering the sources' brightness as the weights. Figure \ref{fig:weighted_std_confidence} illustrates this relationship, with the angular sizes of the sources displayed using colour. There is significantly more scatter in the weighted standard deviation of centroid velocities for sources with lower \fw\ values compared to those with higher \fw\ scores. For instance, the subsamples with lower emission fraction have $\sigma_{v_\text{lsr}}$ in the range of $\sim$5--60 \kms, while those with higher emission fraction have $\sigma_{v_\text{lsr}}$ between 0 and $\sim$30 \kms. Therefore, the negative slope shows that higher \fw\ values correspond to lower weighted standard deviation of cloud centroid velocity, which implies that higher \fw\ values indicate better associations which lead to more stable and consistent velocity measurements hence more reliable in terms of their velocity structure. It is noted that the distribution of sources with relatively low window emission fraction shows a wider spread of values. Notably, sources that are consistent with clouds of increased \fw\ scores show a smoother correlation with $\sigma_{v_\text{lsr}}$. No consistent trend or uniformity is evident regarding the angular sizes (radii) of the sources. This may suggest that either the size of the SMGPS sources does not significantly impact the reliability of their velocity associations or it is possible that the association process is more dependent on other physical properties (e.g. brightness) than size. We also expect this $\sigma_{v_\text{lsr}}$ to change at different Galactic longitudes. The $\ell=342.5$ slice is less than 20 degrees from the Galactic centre, and so the sight-line covers a large column of the Galactic disc, with many spiral arms covering a wide velocity range.

\section{SEDIGISM molecular clouds with extreme virial parameter ($\alpha\geq 10$)}

\begin{table*}
\centering
\caption{Associations of SMGPS sources with SEDIGISM molecular clouds with extreme virial parameter $\alpha_\text{vir}\geq 10$: Columns 1–4 give the SMGPS source name, source iauName(defined based on the galactic coordinates of the source), calculated physical size (radius) of the source, and SMGPS class name. Columns 5–14 give the SEDIGISM cloud name, centroid velocity for the associated SEDIGISM molecular cloud, velocity range of best-matching velocity window where the associated cloud falls, cloud's velocity dispersion, high mass star formation (HMSF), cloud's mass, cloud's average gas surface density, cloud's virial parameter, cloud star formation efficiency (SFE), cloud dense gas fraction (DGF), and cloud morphology structure type. }
\resizebox{\hsize}{!}{
\begin{tabular}{cccccccccccccccc}
\hline
\hline
SMGPS & iauName & Radius & Classname & SEDIGISM & $v_\text{lsr}$ & Window's $v_\text{lsr}$ range 
& Linewidth ($\sigma_v$) & HMSF & Mass & $\Sigma$ & $\alpha_\text{vir}$ & SFE & DGF & Cloud structure\\
(Source Name) & - & (pc) & - & (Cloud Name) & (\kms) & (\kms) & (\kms) & - & (M$_\odot$) 
 & (M$_\odot$\pc) & - &L$_\odot$/M$_\odot$  & - & -\\
\hline
S30033 & G342.389+00.125 & 0.05 & HII & SDG342.382+0.1304 & -7.28 & -10.25, -2.25 & 2.047 & 1.0 & 83 & 197 & 21.56 & 2.73 & 0.47 & concentrated \\
S30034 & G342.384+00.123 & 0.06 & HII & SDG342.382+0.1304 & -7.28 & -10.25, -1.75 & 2.047 & 1.0 & 83 & 197 & 21.56 & 2.73 & 0.47 & concentrated \\
S30048 & G341.844-00.304 & 3.83 & SNR & SDG341.847-0.3162 & -30.20 & -34.25, -24.25 & 2.957 & 0.0 & 3259 & 86 & 10.84 & -- & -- & ring \\
S30049 & G341.672-00.390 & 4.36 & HII & SDG341.847-0.3162 & -30.20 & -38.25, -26.25 & 2.957 & 0.0 & 3259 & 86 & 10.84 & -- & -- & ring \\
S30056 & G341.276-00.352 & 3.18 & HII & SDG341.235-0.3637 & -43.87 & -47.00, -41.50 & 1.642 & 0.0 & 315 & 81 & 11.08 & -- & -- & ring \\
S30058 & G341.156-00.354 & 6.20 & HII & SDG341.235-0.3637 & -43.87 & -47.00, -38.25 & 1.642 & 0.0 & 315 & 81 & 11.08 & -- & -- & ring \\
S30107 & G342.438-00.059 & 0.13 & HII & SDG342.434-0.0620 & -6.01 & -9.50, -2.50 & 2.074 & 1.0 & 22 & 123 & 54.34 & 2.10 & 0.49 & elongated \\
S30108 & G342.428-00.051 & 0.10 & HII & SDG342.434-0.0620 & -6.01 & -8.75, -2.75 & 2.074 & 1.0 & 22 & 123 & 54.34 & 2.10 & 0.49 & elongated \\
S30109 & G342.427-00.050 & 0.40 & HII & SDG342.434-0.0620 & -6.01 & -9.50, -1.50 & 2.074 & 1.0 & 22 & 123 & 54.34 & 2.10 & 0.49 & elongated \\
S30110 & G342.355-00.052 & 0.60 & HII & SDG342.347+0.0118 & -6.20 & -8.25, -3.00 & 1.303 & 0.0 & 39 & 131 & 15.63 & 0.76 & 0.70 & ring \\
S30112 & G342.341+00.003 & 0.10 & unclassified & SDG342.347+0.0118 & -6.20 & -8.25, -3.50 & 1.303 & 0.0 & 39 & 131 & 15.63 & 0.76 & 0.70 & ring \\
S30115 & G341.803-00.430 & 2.25 & unclassified & SDG341.847-0.3162 & -30.20 & -31.25, -29.75 & 2.957 & 0.0 & 3259 & 86 & 10.84 & -- & -- & ring \\
S30122 & G341.985-00.449 & 0.81 & HII & SDG341.962-0.4546 & -11.78 & -14.25, -8.75 & 1.155 & 0.0 & 27 & 78 & 18.95 & 1.54 & 0.56 & clumped \\
S4130 & G342.958-00.013 & 0.07 & unclassified & SDG342.970-0.0172 & -8.45 & -9.25, -6.75 & 1.004 & 1.0 & 30 & 66 & 14.97 & 0.29 & 0.12 & ring \\
S7855 & G341.968+00.225 & 0.36 & HII & SDG341.970+0.2048 & -11.06 & -15.75, -7.50 & 3.692 & 1.0 & 335 & 66 & 60.15 & 1.86 & 0.07 & clumped \\
S7862 & G341.976+00.246 & 0.11 & unclassified & SDG341.970+0.2048 & -11.06 & -14.50, -11.50 & 3.692 & 1.0 & 335 & 66 & 60.15 & 1.86 & 0.07 & clumped \\
S9582 & G341.502-00.424 & 0.85 & unclassified & SDG341.521-0.4463 & -31.60 & -31.75, -28.50 & 1.668 & 0.0 & 118 & 58 & 22.10 & -- & -- & elongated \\

\hline\hline
\end{tabular}
}
\label{tab:extreme alpha cases}
\end{table*}

\section{Merged catalogue table format}
\begin{table*}
\centering
\caption{Format of the SMGPS-SEDIGISM merged catalogue table for the associated complexes.}

\begin{tabular}{c|c|p{12cm}}
\hline
\hline
Name & Unit & Description \\
\hline
source\_name & - & Interacting SMGPS extended source name \\
iauName & - & SMGPS extended source name in IAU format (from {\citealt{2025A&A...Bordiu}})\\
npix & - & Number of pixels in island (from {\citealt{2025A&A...Bordiu}})\\
nested\_ext & - & Child (nested) extended sources (from {\citealt{2025A&A...Bordiu}})\\
morph\_type & - & Morphology tag of the sources e.g. 1=Compact, 2=Point-like, 3=Extended, 5=Diffuse (from {\citealt{2025A&A...Bordiu}})\\
$\ell_s$ & degree & SMGPS source centroid position: Galactic Longitude coordinate (from {\citealt{2025A&A...Bordiu}})\\
$b_s$ & degree & SMGPS source centroid position: Galactic Latitude coordinate (from {\citealt{2025A&A...Bordiu}})\\
radius & - & SMGPS source radius in pixels (from {\citealt{2025A&A...Bordiu}})\\
radius\_wcs & arcmin & SMGPS source angular size (radius) (from {\citealt{2025A&A...Bordiu}})\\
radius\_pc & parsec & SMGPS source calculated physical distance (radius)\\
flux & Jy & SMGPS source measured flux density with background and nested compact sources subtracted (from {\citealt{2025A&A...Bordiu}})\\
$L_r$ & \WHz & SMGPS source calculated radio luminosity\\
classname & - & SMGPS source classification label in string format (from {\citealt{2025A&A...Bordiu}})\\
classid & - & SMGPS source classification id e.g. 0=Unclassified, 2=Galaxy, 3=PN, 4=SNR, 5=Bubble, 6=\ion{H}{ii} (from {\citealt{2025A&A...Bordiu}})\\
cloud\_name & - & SEDIGISM molecular cloud name as per the SEDIGISM naming scheme (from \citealt{2021MNRAS.500.3027D})\\
$\ell_c$ & degree & SEDIGISM cloud's centroid position in Galactic longitude coordinate (from \citealt{2021MNRAS.500.3027D})\\
$b_c$ & degree & SEDIGISM cloud's centroid position in Galactic latitude coordinate (from \citealt{2021MNRAS.500.3027D})\\
$v_\text{lsr}$ & \kms & SEDIGISM cloud's systemic and centroid velocity (from \citealt{2021MNRAS.500.3027D})\\
velocity\_range & \kms & Window velocity range for the best-matching velocity window where all the associated clouds fall\\
pixel\_coverage & - & Fraction of a given SMGPS source covered by SEDIGISM cloud in terms of pixels\\
linewidth ($\sigma_\text{v}$) & \kms & SEDIGISM cloud's velocity dispersion (from \citealt{2021MNRAS.500.3027D})\\
HMSF & - & Cloud High mass star formation (from \citealt{2021MNRAS.500.3027D})\\
dist\_kpc & kiloparsec & Cloud final adopted physical distance (updated by {\citealt{2021A&A...Colombo}})\\
Mass & solMass (M$_\odot$) & SEDIGISM cloud mass (updated by {\citealt{2021A&A...Colombo}})\\
R\_dec & parsec & Cloud deconvolved equivalent radius (updated by {\citealt{2021A&A...Colombo}})\\
column\_density ($N_{\mathrm{H}_2}$) & $cm^{-2}$ & SEDIGISM cloud average column density (updated by {\citealt{2021A&A...Colombo}})\\
surf\_density ($\Sigma$) & M$_\odot$\pc & Cloud deconvolved average gas surface density (updated by {\citealt{2021A&A...Colombo}})\\
$\alpha_\mathrm{vir}$ & - & Cloud deconvolved virial parameter (updated by {\citealt{2021A&A...Colombo}})\\
SFE &L$_\odot$/M$_\odot$ & Cloud star formation efficiency (from \citealt{2021MNRAS.500.3050U})\\
DGF & - & Cloud dense gas fraction (from \citealt{2021MNRAS.500.3050U})\\
by\_eye\_structure & - & Cloud morphology structure type e.g. elongated, ring, clumped, concentrated (from {\citealt{2022A&A...663A..56N}})\\
emission\_fraction ($f_W$) & - & The ratio
of the total integrated intensity within the best emission window to the total integrated intensity of all emission windows in an extracted spectrum\\ 

\hline
\end{tabular}

\label{tab: merged_catalogue}
\end{table*}


\bsp	
\label{lastpage}
\end{document}